\documentclass[12pt]{nyuthesis}
\usepackage{lgrind}
\usepackage{cmap}
\usepackage[T1]{fontenc}
\usepackage{times}
\usepackage{microtype}
\usepackage{epsdice}
\usepackage{bbm}
\usepackage{mathtools}
\usepackage{amsmath}
\usepackage[hyphens]{url}
\usepackage{amssymb}
\usepackage{float}
\usepackage{graphicx}
\usepackage{url}
\usepackage{float}
\usepackage{algorithmicx}
\usepackage[noend]{algpseudocode}
\usepackage{verbatim}
\usepackage{afterpage}
\usepackage{enumitem}
\usepackage{framed}
\usepackage{caption}
\usepackage{subcaption}
\usepackage{fancyhdr}
\usepackage{color}
\usepackage{nameref}
\usepackage{enumitem}
\usepackage{tikz}
\usepackage{comment}
\usepackage{hyperref}
\usepackage{multicol}
\usepackage{pgf-pie} 
\usepackage{multirow}

\usepackage{tikz}
\usepackage{pgfplots}

\usepackage[hyphens]{url}
\usepackage{hyperref}
\hypersetup{
    colorlinks=true,
    linkcolor=black,
    filecolor=black,
    urlcolor=black,
    citecolor=black
}

\usepackage{indentfirst}

\usepackage{amsfonts}

\usepackage{xspace}

\usepackage{microtype}
\usepackage{color}
\usepackage{mathptmx}
\usepackage{times}
\usepackage{amsmath}
\usepackage{amssymb}
\usepackage{bbm}

\usepackage{amsthm}
\usepackage{comment}
\usepackage{algorithm}
\usepackage{algpseudocode}
\usepackage{multirow}
\usepackage{enumitem}
\usepackage{wrapfig}
\usepackage{ctable}
\usepackage{hhline}
\usepackage{soul}
\usepackage[mathcal]{eucal}
\usepackage{lmodern}
\usepackage{tabstackengine}
\usepackage{fancyhdr,appendix,psfrag,layout}

\newtheoremstyle{indented}
  {3pt}
  {3pt}
  {}
  {\parindent}
  {\itshape}
  {.}
  {.5em}
  {}
\makeatother

\theoremstyle{indented}

\usepackage{multirow}

\usepackage{listings}

\usepackage[numbers,sort]{natbib}

\usepackage[all,cmtip]{xy}

\renewcommand{\paragraph}[1]{\vspace{0.1cm}\noindent \textbf{#1}}

\newcommand{\hidecomment}[1]{}
\newcommand{\savespace}[1]{}

\newcommand{\Exp}[0]{\mathbb{E}}

\newcommand{\RR}[0]{\mathbb{R}}

\newcommand{\A}[0]{\mathcal{A}}
\newcommand{\DD}[0]{\mathcal{D}}
\newcommand{\E}[0]{\mathcal{E}}
\newcommand{\G}[0]{\mathcal{G}}
\newcommand{\N}[0]{\mathcal{N}}

\newcommand{\T}[0]{\mathcal{T}}

\newcommand{\matr}[1]{\mathbf{#1}}
\newcommand{\mW}[0]{\matr{W}}

\newcommand{\vect}[1]{\mathbf{#1}}
\newcommand{\vb}[0]{\vect{b}}
\newcommand{\vc}[0]{\vect{c}}
\newcommand{\ve}[0]{\vect{e}}
\newcommand{\vh}[0]{\vect{h}}
\newcommand{\vv}[0]{\vect{v}}

\DeclareMathOperator*{\argmax}{\arg \max}

\newcommand\Chapter[2]{
  \chapter[#1: {\itshape#2}]{#1\\[2ex]\Large#2}
}

\definecolor{g-blue}{HTML}{2E86C1}
\definecolor{g-red}{HTML}{B03A2E}
\definecolor{g-purple}{HTML}{AF7AC5}

\algnewcommand{\And}{\textbf{~and~}}

\newfloat{program}{t}{program}

\fancyheadoffset{0cm}

\fancypagestyle{plain}{%
   \fancyhf{}%
   \fancyhead[R]{\thepage}%
}

\makeatletter
\renewcommand{\l@section}{\@dottedtocline{1}{1.5em}{2.6em}}
\renewcommand{\l@subsection}{\@dottedtocline{2}{4.0em}{3.6em}}
\renewcommand{\l@subsubsection}{\@dottedtocline{3}{7.4em}{4.5em}}
\renewcommand*\l@figure{\@dottedtocline{1}{1em}{3em}}
\makeatother

\begin{document}

\includecomment{covermatter}

\begin{covermatter}
%
%
%

\title{Learning Representations and Agents for Information Retrieval}
\author{Rodrigo Frassetto Nogueira}

\nyunnumber{\underline{N10443173}}
\nyunetid{\underline{rfn216}}

\department{Department of Computer Science and Engineering}

\degree{DOCTOR OF PHILOSOPHY}
\major{Computer Science}

\degreemonth{September}
\degreeyear{2019}
\thesisdate{September 2019}


\supervisor{Kyunghyun Cho}{Prof. Dr.}

\chairman{Name}{Chairman, Department}

\makecover
\cleardoublepage
\maketitle
\cleardoublepage








	\pagenumbering{roman}
	\setcounter{page}{2}
	\makesignaturepage
	\cleardoublepage
	
	\section*{Microfilm/Publishing}

Microfilm or copies of this dissertation may be obtained from:

\singlespace
UMI Dissertation Publishing 

ProQuest CSA 

789 E. Eisenhower Parkway 

P.O. Box 1346 

Ann Arbor, MI 48106-1346

\doublespace

	\section*{Vita}

\onehalfspacing

\noindent \textbf{\large Rodrigo Frassetto Nogueira} 

\subsection*{Education}
\noindent  \textbf{PhD in Computer Science} \hfill Sep. 2014 - Sep. 2019\\
New York University, Tandon School of Engineering \\

\noindent  \textbf{M.S. in Computer Engineering} \hfill Aug. 2013 - Aug. 2014\\
Universidade Estadual de Campinas, FEEC \\

\noindent  \textbf{B.S. in Electrical Engineering} \hfill Jan. 2005 - Dec. 2009\\
Universidade Estadual de Campinas, FEEC \\

\bigskip


\subsection*{Research and Funding}
This research was performed at the NYU Machine Learning for Language (ML$^2$) Lab. Funding for tuition and personal expenses for the first four years was granted by the CAPES Science without Borders scholarship. Funding for the fifth year and traveling to present the papers that originated this work was provided by Prof. Kyunghyun Cho.

\subsection*{Professional Experience}
Rodrigo has an Ms.C. degree from Universidade Estadual de Campinas (UNICAMP), where he developed with prof. Roberto Alencar Lotufo an award-winning algorithm for real vs. fake fingerprint detection. Before that, he worked for Siemens as a software engineer for five years. During that time, he deployed SCADA and Smart Grid systems and was the main inventor of an automated testing equipment for low-voltage control and protection cubicles.

\section*{Dedication}

I dedicate this work to my family and friends, specially my wife, Andrea, my brother, Danilo, my mother, Fatima, my Father, Paulo, and my aunt, Elza.


	\begin{abstractpage}
	A goal shared by artificial intelligence and information retrieval is to create an oracle, that is, a machine that can answer our questions, no matter how difficult they are. 
A more limited, but still instrumental, version of this oracle is a question-answering system, in which an open-ended question is given to the machine, and an answer is produced based on the knowledge it has access to. Such systems already exist and are increasingly capable of answering complicated questions~\cite{seo2016bidirectional,clark2017simple,devlin2018bert}. This progress can be partially attributed to the recent success of machine learning and to the efficient methods for storing and retrieving information, most notably through web search engines.

One can imagine that this general-purpose question-answering system can be built as a billion-parameters neural network trained end-to-end with a large number of pairs of questions and answers. We argue, however, that although this approach has been very successful for tasks such as machine translation, storing the world's knowledge as parameters of a learning machine can be very hard. A more efficient way is to train an artificial agent on how to use an external retrieval system to collect relevant information. This agent can leverage the effort that has been put into designing and running efficient storage and retrieval systems by learning how to best utilize them to accomplish a task.

In this thesis, we present two instances of such an agent. One is constructed to navigate a web of documents, searching for an answer to a given question. This agent makes use of the enormous human work put into curating documents and their links, in particular, the Wikipedia corpus. The second agent takes advantage of existing search engines by rewriting questions to retrieve more relevant answers. Its advantage lies in the use of reinforcement learning, which requires minimal work in specifying a mechanism to rewrite questions that will return more accurate answers. For both agents, we showed that their performance can be higher than strong baselines.

Furthermore, we look inside the search engine --- previously treated as a black box --- and introduce two novel components to improve it. One is a ranking model that uses unsupervised pretraining to re-rank documents more effectively. The other improves the inverted index representation by augmenting documents with predictions of questions that they might correctly answer. We show that these two methods combined can double the retrieval effectiveness of an off-the-shelf search engine.

In a parallel with computer vision, when the AlexNet model~\cite{krizhevsky2012imagenet} almost reduced the error rate by half in an object detection task, it created a revolution that made possible a multitude of applications, such as self-driving cars.
Similarly, with more effective retrieval mechanisms, we aspire that future question answering systems will have a closer resemblance to a research assistant that helps us expand our understanding of the world.
	\end{abstractpage}

\renewcommand{\contentsname}{Table of Contents}

\tableofcontents
\newpage
\listoffigures
\newpage
\listoftables

\end{covermatter}

\pagenumbering{arabic}

\chapter{Introduction}
\label{chap:introduction}

\section{Motivation}
\label{intro-motivation}

Since ancient times, humans dream of omniscient entities that could answer their most urgent doubts. For example, \textit{Pythia}, or the Oracle of Delphi, was consulted about important decisions throughout the ancient classical world. Diviners in ancient China would carve into ox and turtle bones questions regarding future weather, crop planting, military endeavors, and other topics, and the answers would emerge in the form of cracks.\footnote{\url{https://en.wikipedia.org/wiki/Oracle_bone}} Although somewhat appealing, these were, however, imprecise sources of knowledge and unreliable forecasters. Perhaps more accurate but costly ways of obtaining answers involved searches in books, consultations with the elderly, or through experimentation, when possible.

In recent years, with the advent of the Internet, the dream of having access to an oracle has come closer to reality. For example, one can instantly find the most effective methods to avoid sea-sickness, tips to summer vacations, or what a legal expert says on exporting perishable goods. Two decades ago, to obtain this kind of information one had to, at least, visit a specialist or a library.

A key player in this revolution is the search engine: A machine that can find the information we need through text inputs. It has become so popular that one of the largest commercial search engines serves trillions of such requests per year.\footnote{
Source: \url{https://searchengineland.com/google-now-handles-2-999-trillion-searches-per-year-250247} (accessed on 05/26/2018).}

\begin{figure}
\begin{center}
\centerline{\includegraphics[width=0.9\columnwidth]{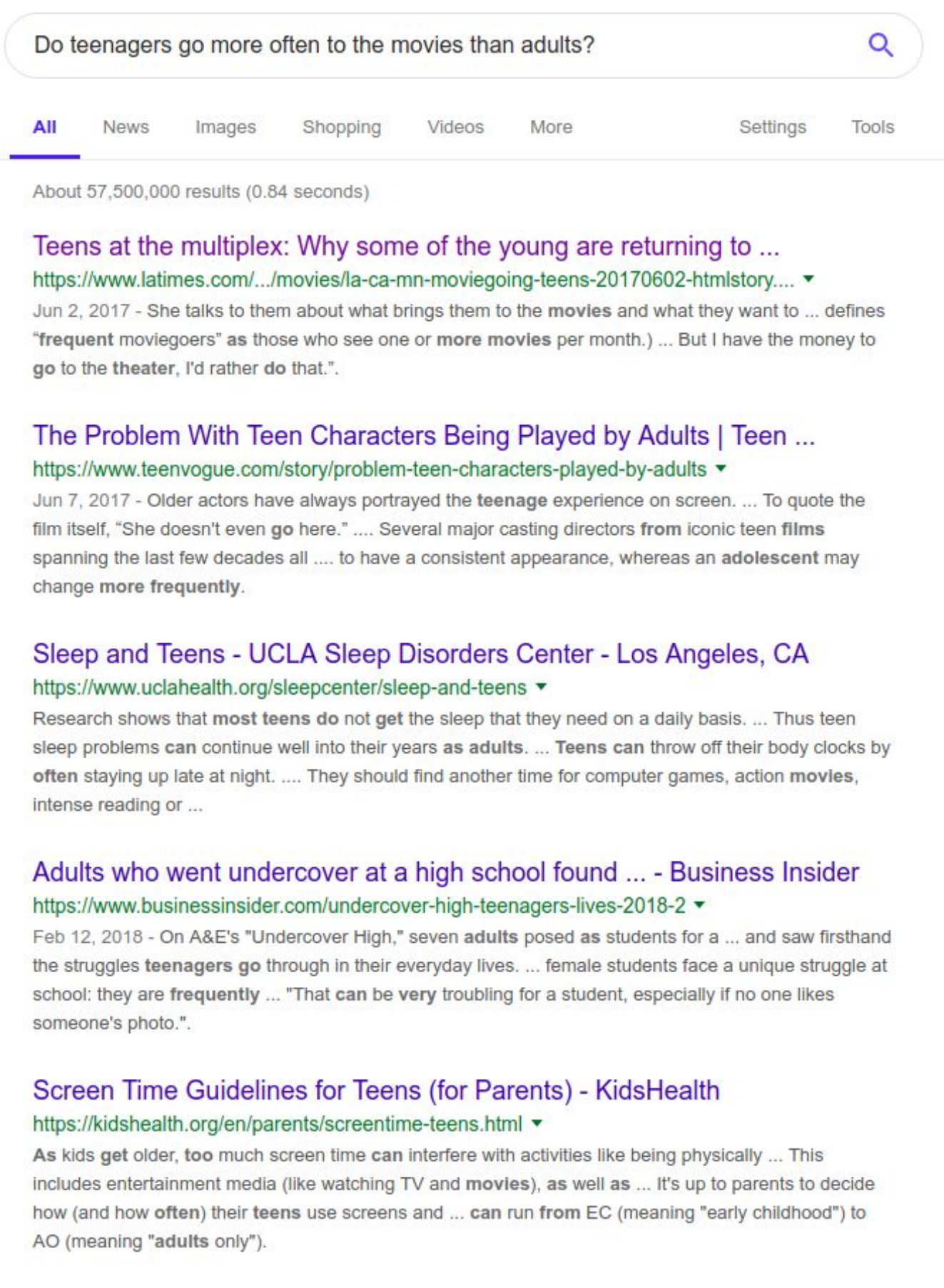}}
\caption{Top-5 results returned by a popular search engine when asked: ``Do teenagers go more often to the movies than adults?'' (as of 24/05/2019). The first result is partially relevant, whereas the others are irrelevant.
}
\label{fig:search_results}
\end{center}
\end{figure}

Despite being very useful, machines still cannot successfully answer many questions. For instance, when asked, ``Do teenagers go more often to the movies than adults?'', the top document returned by a popular search engine contains only part of the answer (i.e., the absolute number of teenagers going to the movies in the U.S. in 2017), and the remaining results are irrelevant (Figure~\ref{fig:search_results}).
To obtain a complete answer, the user might need to rephrase or break the query, each part addressing one aspect of the original question. In our example, the user might need to independently issue the following queries: ``Number of movie tickets sold by age in 2018'', ``Demographics of moviegoers'', ``Number of movie tickets sold by age in the U.S. (or China, or Europe)'', etc. After collecting the results, the user might still need to aggregate them in a spreadsheet in order to have a complete answer. In other words, despite using a state-of-the-art search engine, the user still had to go through a slow and laborious process of obtaining the answer to the original question.

One of the main reasons for this low retrieval effectiveness is the reliance of most existing search engines on keyword match to retrieve the initial set of documents. 
In such cases, a relevant document will not be retrieved if its terms do not match the ones in the query.
This commonly known as the ``vocabulary mismatch'' problem, and users often spend a reasonable amount of time getting familiarized with the terms that are commonly used in the field of interest. 
An analogy is that of a tourist visiting a new city and needs to go from one place to another by asking for directions. She needs to learn the names of the main streets and landmarks to facilitate her navigation through this unknown environment. Likewise, a web user needs to learn the terms used in the relevant web pages or documents to find the desired information.
This problem is aggravated in specialized areas like medicine or law, in which a user might experience an avalanche of new terms when she explores areas just outside her main field of expertise.

To further support our claim that existing retrieval mechanisms do not work well at least in a handful of use cases, we can quantify their effectiveness. One straightforward way is to download publicly available academic question-answering datasets and compute the percentage of questions that can be answered based solely on documents retrieved by existing search engines. For example, on the Natural Questions dataset~\cite{47761}, only half of its questions had an answer in the top returned Wikipedia article. Similarly, in the MS MARCO passage retrieval task~\cite{nguyen2016ms}, the popular BM25 algorithm~\cite{robertson1995okapi} retrieves a relevant paragraph among the top-10 for only 40\% of the questions. In another three datasets (TREC-CAR~\cite{dietz2017trec}, Jeopardy~\cite{nogueira2016end}, and MSA~\cite{nogueira2017task}), among the top-1000 documents retrieved by BM25, correct documents are present only for 30\%-60\% of the queries. In addition, \citet{chen2017reading} found that the performance of their question-answering system drops from 78\% F1, when the correct document is given, to 30\% when the retrieval task is taken into account.

Two of the datasets used in this brief analysis, MS MARCO and Natural Questions, have queries issued by real users to commercial search engines, i.e., Bing and Google, respectively. Although this data contains daily information needs of real users, a user's true information need is only partially represented. The reason is that many users have learned what these machines can and cannot answer. For example, a user might decide to not even issue the query ``Do teenagers go more often to the movies than adults?'' because she anticipates that the machine will probably not return a concise and correct answer. This limitation of the current technology, therefore, hides more complex information needs. In other words, if we have access to more effective answering machines, we will probably start asking more complicated questions.

\section{Contributions}
\label{sec:intro-contributions}

At this point, we hope to have convinced the reader that we need better retrieval mechanisms if we aim at answering more complicated questions. This is precisely the goal of this thesis, whose proposed methods are summarized next.

\paragraph{Retrieval as Web Navigation (Chapter~\ref{chap:webnav})} Here, we do not aim to improve existing search engines by modifying or replacing their components.
Instead, we propose a novel search mechanism in its entirety:\ An agent that searches for relevant information in a corpus by following the hyperlinks that connect the documents.
Because this search mechanism does not rely on keyword match to perform retrieval, the vocabulary mismatch problem is mitigated.
The agent is a neural network trained with supervised learning and fine-tuned with reinforcement learning. For the supervised training, the ground truth navigation paths are obtained using a shortest path algorithm. 
In the reinforcement finetuning step, the reward is a binary function that tells the agent if the correct document for a given question has been reached.

We evaluate our proposed navigational agent on a challenging dataset of \textit{Jeopardy!} questions and show that it performs better than traditional index-based search engines in the densely-linked Wikipedia corpus.

\paragraph{Retrieving from a Black Box (Chapter~\ref{chap:query_reformulation})} We leverage the capabilities of modern search engines through an agent that learns how to use them to obtain better answers. The underlying search engine is treated as a black box, i.e., its internal mechanism is hidden from our agent, so the same method can be used to optimize different types of systems as long as they share a common interface, namely, text as input and output.

A key characteristic of this work is the use of reinforcement learning (RL) to train the agent in reformulating queries to increase the chances of retrieving relevant documents. The main advantage of using RL is that the agent directly optimizes the metric of choice (recall, MAP, NDCG, etc.). Alternative methods require manually-defined rules for reformulating queries or training data with pairs of original and reformulated queries.

The good results achieved with this method led us to further extend it into using multiple reformulation agents. In this framework, a set of diverse reformulation agents in addition to an answer-aggregation module resulted in better effectiveness than a standard ensemble counterpart. Also, given the same computational budget, it can be trained in a fraction of the time when compared to the single-agent version due to the ease of parallelization of the proposed method.

\paragraph{Looking into the Black Box (Chapter~\ref{chap:blackbox})} Here, we depart from the black box paradigm of the previous chapters and explore methods to enhance the internal mechanism of the search engine. We show that the effectiveness of the document ranking stage can be largely improved by using a neural model pretrained on a language modeling task in a high-quality corpus (e.g. Wikipedia). This model takes as an input the concatenation of the minimally preprocessed query and document texts, and computes a relevance score. The model learns the interactions between query and document terms though the self-attention mechanism~\cite{vaswani2017attention}.

Subsequently, we introduce a technique to improve the representation of the inverted index. We observe that query reformulation is an algorithm that translates query language into document language. We then invert this direction and translate from the document language to the query language. We achieve this by training with supervised learning an off-the-shelf translation model that takes as an input a document and predicts queries to which the document might be relevant. Once trained, each document in the corpus is expanded with its predicted queries and then indexed using a vanilla inverted indexing algorithm.

We show the effectiveness of these two novel components (i.e., document expansion and re-ranker) by achieving the state of the art in two retrieval tasks.

\section{Organization}

This thesis is organized as follows. We introduce the web navigation agent in Chapter~\ref{chap:webnav} and the query reformulation agent in Chapter~\ref{chap:query_reformulation}. We present the novel search engine components in Chapter~\ref{chap:blackbox} and, finally, our conclusion in Chapter~\ref{chap:conclusion}.

\Chapter{A Search Engine from Scratch}{Retrieval as Web Navigation}
\label{chap:webnav}

Over the past five decades, the introduction of digital computers created a revolution on how we access and store information. In particular, digital communication networks gave rise to the Internet, which resulted in the generation and spread of an endless amount of textual, audio, and visual data. Search engines then came to organize this data and provide easy access to it via a simple interface, mostly through short text inputs called queries. 

The way most search engines work is simple to explain.\footnote{For a detailed explanation, see~\citet{larson2010introduction}.}
First, the documents from websites are downloaded by machines called ``web crawlers'' or ``spiders.'' This data is converted into the so-called inverted index. It is, in essence, a dictionary whose keys are the words in the documents and values are pointers to the documents that contain the word. Finally, when a user enters a query, the words in it are matched to the keys in the inverted index, and the corresponding documents are returned, ordered by a ranking algorithm.

Although successful in many domains, index-based search engines have limitations. First, if the terms in a query do not match the ones used in a relevant document despite being semantically the same (i.e., ``auto sellers'' vs. ``I want to buy a car''), the index-lookup might not return the correct document, unless additional mechanisms such as query rewrite~\cite{carpineto2012survey} are employed. Second, if a query is long, like a conversation with a digital assistant, matching all terms in it with the ones in the inverted index will result in a large set of returned documents, and ranking them will be expensive.

In this chapter, we propose a goal-driven web navigation agent as an alternative to index-based search engines. The proposed goal-driven web navigation environment consists of the whole website as a graph, in which the web pages are nodes and hyperlinks are directed edges. An agent is given a query and navigates the network, starting from a predefined node, to find a target node that contains the answer to the query. 
 
We release a software tool, called WebNav, that converts a given website into a goal-driven web navigation task. As an example of its use, we provide WikiNav, which was built from English Wikipedia. We design artificial agents based on neural networks (called NeuAgents) trained with supervised and reinforcement learning, and report their respective performances on the task as well as the performance of human volunteers.

Furthermore, we extend the WikiNav with an additional set of queries that are constructed from {\it Jeopardy!} questions, to which we refer by WikiNav-Jeopardy. We evaluate the proposed NeuAgents against the three search-based strategies; (1) SimpleSearch, (2) Apache Lucene with BM25~\cite{robertson1995okapi} ranking function, and (3) Google Search API. The result indicates that the NeuAgents outperform those index-based search engines, implying a potential for the proposed task as a good proxy for practical applications such as document retrieval, question answering and focused crawling.

\section{Goal-driven Web Navigation}
\label{sec:task}

A task $\T$ of goal-driven web navigation is characterized by 
\begin{align}
    \label{eq:task}
    \T=(\A, s_S, \G, q, R, \Omega).
\end{align}

The world in which an agent $\A$ navigates is represented as a graph $\G=(\N,\E)$.
The graph consists of a set of nodes $\N=\left\{ s_i \right\}_{i=1}^{N_{\N}}$ 
and a set of directed edges $\E=\left\{e_{i,j} \right\}$ connecting those nodes.
Each node represents a page of the website, which, in turn, is represented
by the natural language text $\DD(s_i)$ in it. There exists an edge going from a page $s_i$ to $s_j$ if and only if there is a hyperlink in $\DD(s_i)$ that points to $s_j$. One of the nodes is designated as a starting node $s_S$ from which any
navigation begins. A target node is the one whose natural language description
contains a query $q$, and there may be more than one target node.

At each time step, the agent $\A$ {\em reads} the natural language description
$\DD(s_t)$ of the current node in which the agent has landed. At no point, the
whole world, consisting of the nodes and edges, nor its structure or map (graph
structure without any natural language description) is visible to the agent, thus
making this task {\em partially observed} decision-making.

Once the agent $\A$ reads the description $\DD(s_i)$ of the current node $s_i$,
it can take one of the actions available. A set of possible actions is defined
as a union of all the outgoing edges $e_{i,\cdot}$ and the {\em stop} action,
thus making the agent have {\em state-dependent} action space.

Each edge $e_{i,k}$ corresponds to the agent jumping to a next node $s_k$, while
the stop action corresponds to the agent declaring that the current node $s_i$ 
is one of the target nodes.  Each edge $e_{i,k}$ is represented by the description
of the following node $\DD(s_k)$. In other words, deciding which action to take is
equivalent to taking a peek at each neighboring node and seeing whether that
node is likely to ultimately lead to a target node. 

The agent $\A$ receives a reward $R(s_i, q)$ when it chooses the stop action.
This task uses a simple binary reward, where 
\begin{equation}
    R(s_i, q) = \left\{ \begin{array}{l l}
            1,&\text{if }q \subseteq \DD(s_i) \\
            0,&\text{otherwise}
        \end{array}\right.
\end{equation}

\begin{figure}
    \centering
    \includegraphics[width=0.99\columnwidth]{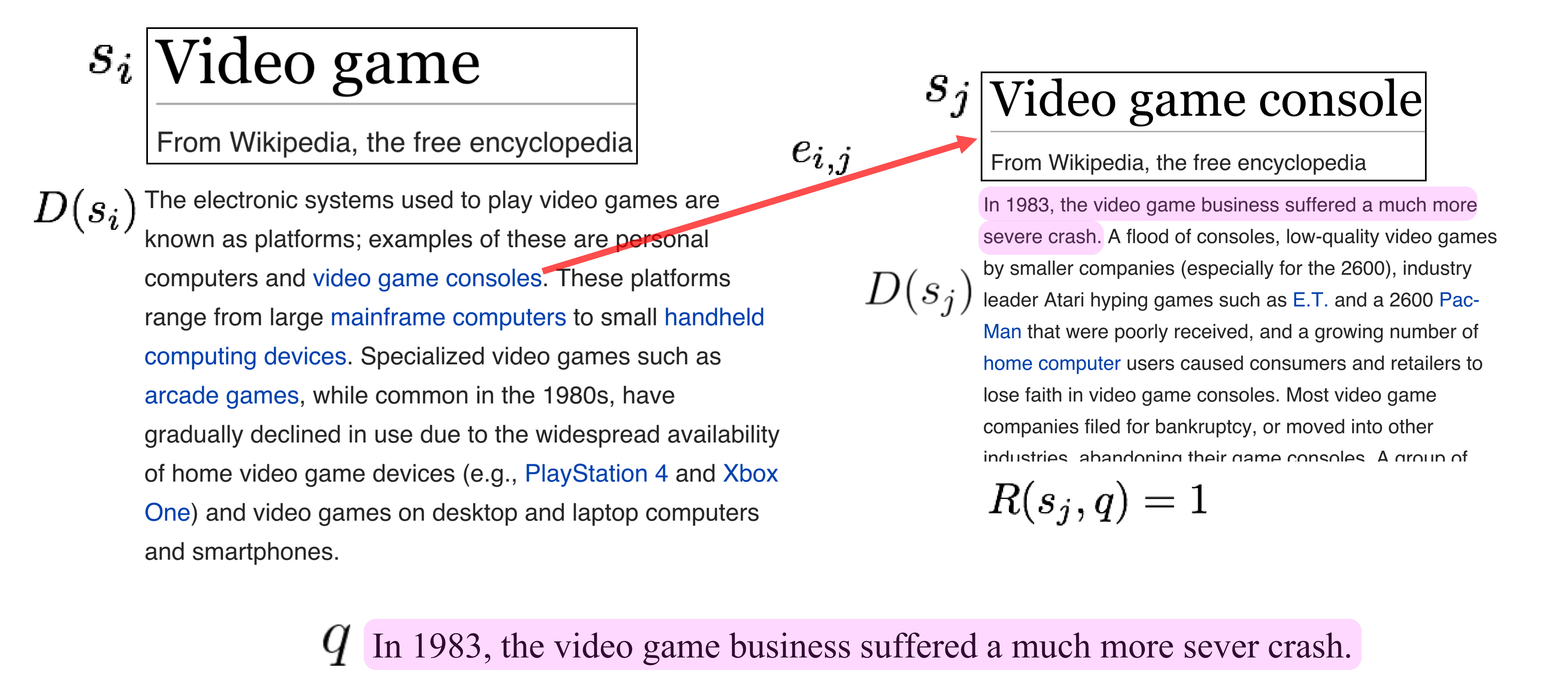}
    \caption{Graphical illustration of a world in the proposed goal-driven
        web navigation. We show two nodes $s_i$ and $s_j$ with their
        contents $\DD(s_i)$ and $\DD(s_j)$, respectively. 
        The reward $R$ is $1$ if and only if a node includes a query
        sentence $q$. In this case, the reward at the node $s_j$, i.e.,
    $R(s_j, q)$ is 1, but $R(s_i, q)=0$.}
    \label{fig:world}

\end{figure}

See Figure~\ref{fig:world} for a graphical illustration of this world.

\paragraph{Constraints:}
It is clear that there exists a policy for the agent to succeed at every trial, which is to traverse the graph breadth-first until the agent finds a node in which the query appears. To avoid this kind of degenerate policies, the task includes a set of rules/constraints
$\Omega$. 

More specifically, there are four constraints:
\begin{enumerate}
    \itemsep -.2em
    \item An agent can follow at most $N_n$ edges at each node.
    \item An agent has a finite memory of size smaller than $\T$.
    \item An agent moves up to $N_h$ hops away from $s_S$.
    \item A query of size $N_q$ comes from at least two hops away from the starting node.
\end{enumerate}

The first constraint alone prevents degenerate policies, such as breadth-first search, forcing the agent to make good decisions as possible at each node.
The second one further constraints ensure that the agent does not cheat by using earlier trials to reconstruct the whole graph structure (during test time) or to store the whole world in its memory (during training.) The third constraint, which is optional, is there for computational consideration.
The fourth constraint is included because the agent is allowed to read the content of the next node.

\subsection{Controlled Levels of Difficulty}

Three main control parameters affect the difficulty of the task. They are the maximum number of explored edges per node $N_n$, the maximum number of allowed hops $N_h$ and the size of each query $N_q$. 


\paragraph{Maximum number of explored edges per node $N_n$:} 
If this parameter is high, an agent does not have to be confident about correct
actions at each node, as there is a possibility of exploring other outgoing
edges. If $N_n$ goes to infinity, the agent can simply perform
breadth-first search. If $N_n=1$, the agent must select the correct outgoing
edge at each node, otherwise it fails the task. We use $N_n=4$.


\paragraph{Maximum number of allowed hops $N_h$:}
This parameter must be selected a priori generating a dataset from an existing website because this is used to select queries. The larger $N_h$, the more difficult the task is. 


\paragraph{Size of query $N_q$:}
The query is effectively the only source of clue the agent can use to plan its
path from the starting node to a target node. Often a longer query contains more
information, leading to easier navigation by the agent. We consider the number
of sentences contained in each query as its size. Later, in the experiments, we
show that it is indeed true that there is a positive correlation between the
size of the query and the difficulty of the task.

\section{WebNav and WikiNav}
\label{sec:webnav}

\subsection{WebNav: Software}

As a part of this work, we build and release a software tool which turns a
website into a goal-driven web navigation task.\footnote{The source code and
datasets are publicly available at \url{https://github.com/nyu-dl/WebNav}.}
We call this tool {\em WebNav}.
Given a starting URL, the WebNav reads the whole website, constructs a graph
with the web pages in the website as nodes. Each node is assigned a unique
identifier $s_i$. The text content of each node $\DD(s_i)$ is a cleaned version of the actual HTML content of the corresponding web page. We provide a cleanup function for Wikipedia, and it is easy to plug in a new cleanup function for another website. The WebNav turns intra-site hyperlinks into a set of edges $e_{i,j}$. 

In addition to transforming a website into a graph $\G$ from
Equation~\eqref{eq:task}, the WebNav automatically extracts sentences as queries from the 
nodes' texts and divides them among training, validation, and test sets.
We ensure that there is no overlap among these sets by making each target node, from which a query is selected, belongs to only one of the three sets.

Each generated example is defined as a tuple
\begin{equation}
    \label{eq:example1}
    X = (q, s^*, p^*)
\end{equation}
where $q$ is a query from a web page $s^*$ which was found following a
randomly selected path $p^*=(s_S, \ldots, s^*)$. In other words, the WebNav
starts from a starting page $s_S$, random-walks the graph for a predefined
number of steps ($N_h/2$, in our case), reaches a target node $s^*$ and selects
a query $q$ from $\DD(s^{*})$. 

A query consists of $N_q$ sentences and is selected among top-5 candidates in
the target node with the highest average TF-IDF, thus discouraging the WebNav
from choosing a trivial query.

For the evaluation purpose alone, it is enough to use only a query $q$ itself as
an example. However, we include both one target node (among potentially many
other target nodes) and one path from the starting node to this target node
(again, among many possible connecting paths) so that they can be exploited when
training an agent. They are not to be used when evaluating a trained agent.


\begin{table}[ht]
    \centering
    \begin{minipage}{0.6\columnwidth}
        \centering
        \begin{tabular}{r || c | c }
            & Hyperlinks & Words \\
            \hline
            \hline
            Avg. & 4.29 & 462.5 \\
            $\sqrt{\text{Var}}$ & 13.85 & 990.2 \\
            Max & 300 & 132881 \\
            Min & 0 & 1 \\
        \end{tabular}
    \end{minipage}
    \hfill
    \begin{minipage}{0.39\columnwidth}
        \caption{Per-page statistics of English Wikipedia used to build WebNav-$n$
        tasks.}
        \label{tab:wiki_stat}
    \end{minipage}
\end{table}

\subsection{WikiNav: A Wikipedia-based Navigation Task}

Using the WebNav, we built a goal-driven navigation task using
Wikipedia as a target website. We used the dump file of the English Wikipedia
from September 2015, which consists of more than five million web pages. We
built a set of separate tasks with different levels of difficulty by varying the
maximum number of allowed hops $N_h \in \left\{ 4, 8, 16 \right\}$ and the size
of query $N_q \in \left\{1, 2, 4 \right\}$. We refer to each task by
\mbox{WikiNav-$N_h$-$N_q$}. 

For each task, we generate training, validation, and test examples from the pages half as many hops away from a starting page as the maximum number of hops allowed.\footnote{
    This limit is an artificial limit we chose for computational reasons. Such
    limitation is not necessary, and the difficulty of the task can be
    arbitrarily increased by choosing a much larger number of hops and selecting
    queries from any page at least two hops away from a starting page.
} 
We use ``Category: Main topic classifications'' as a starting node $s_S$.

As a minimal cleanup procedure, we excluded meta articles whose titles start
with ``Wikipedia.'' Any hyperlink that leads to a web page outside Wikipedia is removed in advance together with the following sections: ``References,''
``External Links,'' ``Bibliography,'' and ``Partial Bibliography.''

In Table~\ref{tab:wiki_stat}, we present basic per-article statistics of the
English Wikipedia. It is evident from these statistics that the world of
\mbox{WikiNav-$N_h$-$N_q$} is large and complicated, even after the cleanup procedure. 

We ended up with a fairly small dataset for \mbox{WikiNav-4-*}, but large for
\mbox{WikiNav-8-*} and \mbox{WikiNav-16-*}. See Table~\ref{tab:stat} for details.

\begin{table}
    \centering
    \small
    \begin{tabular}{r c c c c}
        & {\bf WikiNav-4-*} & {\bf WikiNav-8-*} & {\bf WikiNav-16-*} & {\bf WikiNav-Jeopardy}\\ 
        \toprule
        {\bf Train} & 6.0k & 1M & 12M & 113k \\
        \midrule
        {\bf Valid} & 1k & 20k & 20k & 10k \\
        \midrule
        {\bf Test } & 1k & 20k & 20k & 10k \\
    \end{tabular}
    \caption{Number of examples of WikiNav-4-*,
    \label{tab:stat}
    WikiNav-8-*, WikiNav-16-* and WikiNav-Jeopardy.}
\end{table}

\subsection{WikiNav-Jeopardy: {\it Jeopardy!} on WikiNav}
\label{sec:wikijeo}

One of the potential practical applications utilizing a goal-driven navigation agent is question answering based on world knowledge. 
In this Q\&A task, a query is a question, and an agent navigates a given information network, e.g., a website, to retrieve an answer. 
In this section, we propose and describe an extension of the WikiNav, in which query-target pairs are constructed from actual {\it Jeopardy!} question-answer pairs.
We refer to this extension of WikiNav by {\it WikiNav-Jeopardy}. 

\begin{table}
    \centering
    \begin{tabular}{ll}
        {\bf Query} & {\bf Answer} \\ 
        \toprule
        \small For the last 8 years of his life, Galileo was under &
        \multirow{2}{*}{Copernicus} \\
        \small house arrest for espousing this man's theory. & \\
        \midrule
        \small In the winter of 1971-72, a record 1,122 inches of snow fell &
        \multirow{2}{*}{Washington} \\
        \small at Rainier Paradise Ranger Station in this state. & \\
        \midrule
        \small This company's Accutron watch, introduced in 1960, &
        \multirow{2}{*}{Bulova} \\
        \small had a guarantee of accuracy to within one minute a month. & 
    \end{tabular}
    \caption{Sample query-answer pairs from WikiNav-Jeopardy.}
    \label{tab:JeoQuestions}
\end{table}

We first extract all the question-answer pairs from {\it J! Archive}\footnote{
    \url{http://www.j-archive.com}
}, which has more than 300k such pairs. We keep only those pairs whose answers are titles of Wikipedia articles, leaving us with 133k pairs. We divide those pairs into 113k training, 10k validation, and 10k test examples while carefully ensuring that no article appears in more than one partition. Additionally, we do not shuffle the original pairs to ensure that the train and test examples are from different episodes.

For each training pair, we find one path from the starting node ``Category: Main Topic Classification'' to the target node and include it for supervised learning. For reference, the average number of hops to the target node is 5.8, the standard deviation is 1.2, and the maximum and minimum are 2 and 10, respectively.
See Table~\ref{tab:JeoQuestions} for sample query-answer pairs.

\section{NeuAgent: Neural Network based Agent}
\label{sec:neuagent}

Here, we describe the neural network based agents built with a minimal set of prior knowledge.

\subsection{Core Function}

The core of the NeuAgent is a parametric function $f_{\text{core}}$ that takes
as input the content of the current node $\phi_c (s_i)$ and a query $\phi_q
(q)$, and that returns the hidden state of the agent. This parametric function
$f_{\text{core}}$ can be implemented either as a feedforward neural network
$f_{\text{ff}}$:
\begin{equation}
    \vh_t = f_{\text{ff}}(\phi_c (s_i), \phi_q (q)),
\end{equation}
which does not take into account the previous hidden state of the agent or as a recurrent neural network $f_{\text{rec}}$:
\begin{equation}
    \vh_t = f_{\text{rec}}(\vh_{t-1}, \phi_c (s_i), \phi_q (q)).
\end{equation}
We refer to these two types of agents by {\it NeuAgent-FF} and {\it
NeuAgent-Rec}, respectively. For the NeuAgent-FF, we use a single $\tanh$
layer, while we use long short-term memory (LSTM) units
\cite{hochreiter1997long} for the NeuAgent-Rec.

Based on the new hidden state $\vh_t$, the NeuAgent computes the probability
distribution over all the outgoing edges $e_i$. The probability of each
outgoing edge is proportional to the similarity between the hidden state $\vh_t$
such that
\begin{equation}
    \label{eq:edge_p}
    p(e_{i, j}|\tilde{p}) \propto \exp \left( \phi_c (s_j)^\top \vh_t \right).
\end{equation}
Note that the NeuAgent peeks at the content of the next node $s_j$ by
considering its vector representation $\phi_c (s_j)$.  In addition to all the
outgoing edges, we also allow the agent to {\em stop} with the probability
\begin{equation}
    \label{eq:stop_p}
    p(\emptyset|\tilde{p}) \propto \exp \left( \vv_{\emptyset}^\top \vh_t \right),
\end{equation}
where the stop action vector $\vv_{\emptyset}$ is a trainable parameter. In the case of NeuAgent-Rec, all these (unnormalized) probabilities are conditioned on the history $\tilde{p}$, which is a sequence of actions (nodes) selected by the agent so far.

We divide these unnormalized probabilities by 
\begin{equation}
    Z(\tilde{p}) = \exp \left( \vv_{\emptyset}^\top \vh_t \right) + \sum_{e_{i,j} \in
    e_{i,\cdot}} \exp \left( \phi_c (s_j)^\top \vh_t \right)
\end{equation}
to obtain the probability distribution over all the possible actions at the
current node $s_i$, which is known as {\em softmax} normalization.

The NeuAgent then selects its next action based on this action probability
distribution (Eqs.~\eqref{eq:edge_p} and \eqref{eq:stop_p}). If the stop action
is chosen, the NeuAgent returns the current node as an answer and receives a
reward $R(s_i, q)$, which is one if correct and zero otherwise. If the agent
selects one of the outgoing edges, it moves to the selected node and repeats
this process of {\em reading} and {\em acting}. 

\begin{figure}
    \centering
    \includegraphics[width=.7\columnwidth]{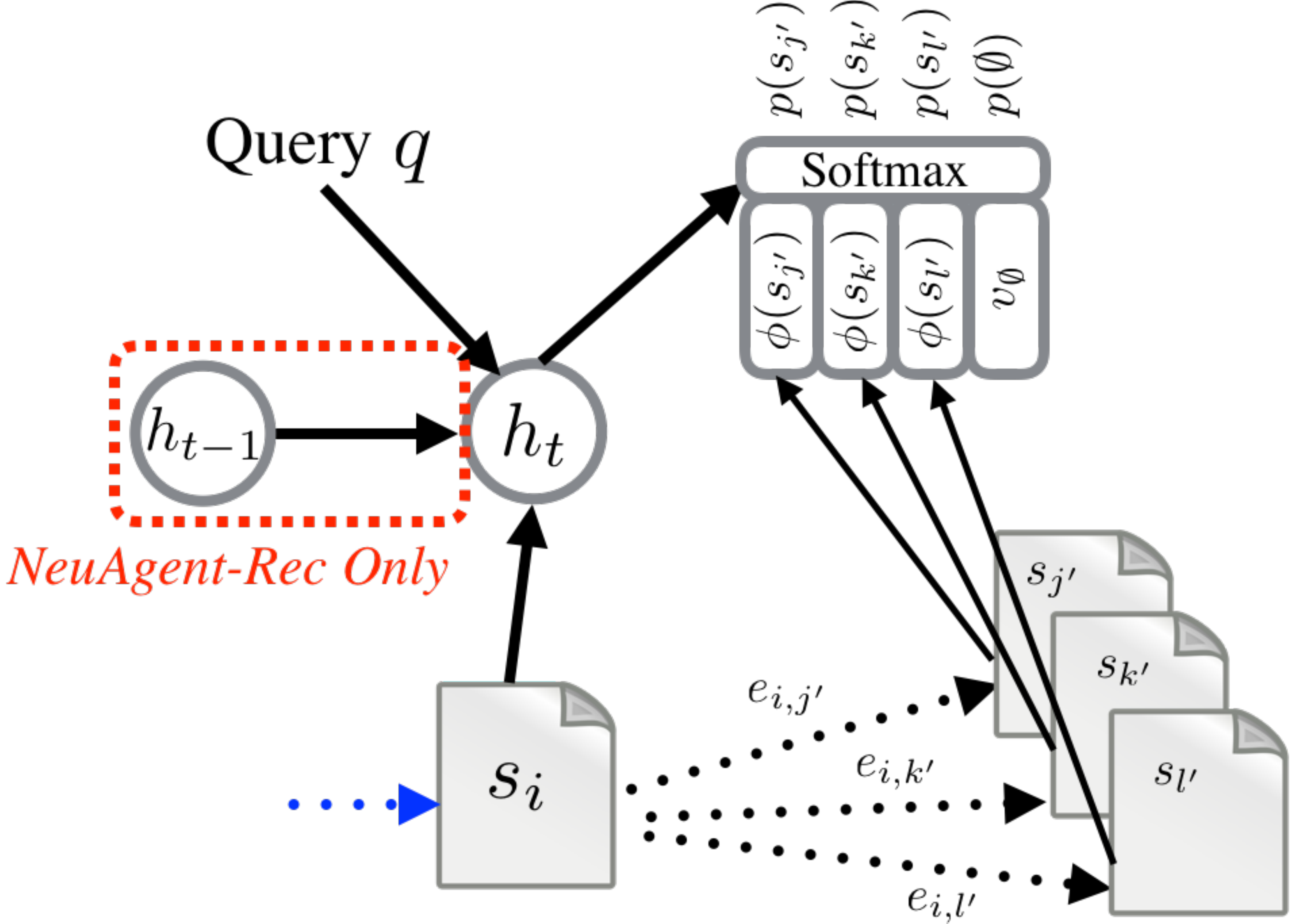}
    \caption{Graphical illustration of a single step performed by the baseline
    model, NeuAgent.}
    \label{fig:agent}
\end{figure}

See Figure~\ref{fig:agent} for a single step of the described NeuAgent.

\subsection{Content Representation}

The NeuAgent represents the content of a node $s_i$ as a vector $\phi_c(s_i) \in
\RR^d$. In this work, we consider two types of representations.
In the first one, the content is represented as the average of the simple
continuous bag-of-words vector (BoW) of each word present in the document:
\begin{equation}
    \phi_c(s_i) = \frac{1}{|\DD(s_i)|} \sum_{k=1}^{|\DD(s_i)|} \ve_k.
\end{equation}
We use word vectors $\ve_k$ from a pretrained continuous bag-of-words model
\cite{mikolov2013efficient}. These word vectors are fixed
and not updated when training the NeuAgent.

In the other type of representation, we make use of the attention
mechanism~\cite{bahdanau2014neural} to form the representation vector of the
document.
A document consists of a set of sections $\{\text{sec}_j(s_i)\}_{j=1}^M$, each
represented as the average continuous bag-of-words vector of
the words in it:
\begin{equation}
\vb_j = \sum_{k=1}^{|\text{sec}_j(s_i)|} \ve_k
\end{equation}

Each vector $\vb_j$ is convoluted by a trainable 1-D weight matrix
$\mW \in \RR^{u \times d}$ to form the context vector $\vc_j$ of the section:

\begin{equation}
\vc_j = \sum_{j'=j-\frac{u}{2}}^{j+\frac{u}{2}} \mW_{j'} \vb_{j'}
\end{equation}
where $u$ is the window size.
A score $\beta_j^t$ is computed for each context vector using a parametric
function, such as a feedforward neural network, that takes as inputs the query
embeddings, the previous hidden state, and the context $\vc_j$:
\begin{equation}
\beta_j^t = f_{\text{doc}}(\phi_q(q), \vh_{t-1}, \vc_j)
\end{equation}

The scores are made to sum to 1 through a {\em softmax} function \cite{Bridle1990}:
\begin{equation}
\alpha_j^t = \frac{\exp(\beta_j^t)}{\sum_{l=1}^M \exp(\beta_l^t)}
\end{equation}
and the vector representation of the document
at step $t$ is obtained as the weighted sum of the context vectors:
\begin{equation}
\phi_c(s_i) = \frac{1}{M} \sum_{j=1}^M \alpha_j^t \vc_j
\end{equation}

\subsection{Query Representation}

We consider the same two types of representation for a query $q$.
For the BoW representation, the vector is formed as the average of the word
vectors of each word present in the query:
\begin{equation}
    \phi_q(q) = \frac{1}{|q|} \sum_{k=1}^{|q|} \ve_k.
\end{equation}
The second type is the attention-based representation. The query
is projected to a context vector $\vc_k$, formed by the word embedding $\ve_k$
convoluted by a 1-D trainable weight matrix
$\mW \in \RR^{u \times d}$:
\begin{equation}
\vc_k = \sum_{k'=k-\frac{u}{2}}^{k+\frac{u}{2}} \mW_{k'} \ve_{k'}
\end{equation}
and the scores are computed by another parametric function (a feedforward neural
network, in this work):
\begin{equation}
\beta_k^t = f_{\text{query}}(\vh_{t-1}, \vc_k)
\end{equation}

The scores are made to sum to 1 through a {\em softmax} function and the 
vector representation at step $t$ is obtained as the weighted sum of
the context vectors:
\begin{equation}
\alpha_k^t = \frac{\exp(\beta_k^t)}{\sum_{l=1}^{|q|} \exp(\beta_l^t)}
\end{equation}
\begin{equation}
\phi_q(q) = \frac{1}{|q|} \sum_{k=1}^{|q|} \alpha_k^t \vc_k
\end{equation}

We empirically compare these two types of representations for both
content and query in section~\ref{sec:wikiquant}.

\subsection{Inference: Beam Search}
\label{sec:beamsearch}

Once the NeuAgent is trained, there are a number of approaches to using it for solving the navigation task.
The most naive approach is to let the agent make a greedy decision at each time step, i.e., following the outgoing edge with the highest probability $\argmax_{k} \log p(e_{i,k}|\ldots)$.
A better approach is to exploit the fact that the agent is allowed to explore up to $N_n$ outgoing edges per node.
This naturally leads to approximate decoding, and we use a simple, forward-only beam search with the beam width capped at $N_n$. 
The beam search keeps the $N_n$ most likely traces, in terms of $\log p(e_{i, k} |\ldots)$, at each time step.

\subsection{Training Strategies: Supervised Learning}

In this work, one of the training strategies we investigate is supervised
learning, in which we train the agent to
follow an example trace $p^*=(s_S, \ldots, s^*)$ included in the training set at
each step (see Equation~\eqref{eq:example1}). In this case, for each training
example, the training cost is
\begin{equation}
    \label{eq:c_sup}
    C_{\text{sup}} = -\log p(\emptyset|p^*) - \sum_{k=1}^{|p^*|} \log p(p^*_k |
    p^*_{<k}).
\end{equation}
This per-example training cost is fully differentiable with respect to all the
parameters of the neural network, and we use stochastic gradient
descent (SGD) algorithm to minimize this cost over the whole training set:
\begin{equation}
    \theta \leftarrow \theta - \eta \nabla C_{\text{sup}},
\end{equation}
where $\theta$ is a set of all the parameters, and $\nabla C_{\text{sup}}$ is
the gradient which can be efficiently computed by backpropagation
\cite{rumelhart1986learning}. This allows the entire model to be trained in an
end-to-end fashion, in which the query-to-target performance is optimized
directly.

This approach of supervised learning for structured output prediction\footnote{
    We can consider any problem of sequential decision making as a structured output prediction where the structured output space contains sequences of actions.
} is known to be prone to accumulating errors as it solves the task
\cite{ross2010efficient}. This is mainly because the agent never sees a wrong node during training, and it tends to fail more easily when it ends up in an unseen node at test time. 

\subparagraph{Entropy Regularization:}
We observed that the action distribution in Equation~\eqref{eq:edge_p} was highly
peaked when the agent was trained with supervised learning. This phenomenon led
to the trained agent not being able to exploit the advantage of beam search
during test time. We address this issue by regularizing the negative entropy of
the action distribution. This is done by adding the following regularization
term to the original cost function in Equation~\eqref{eq:c_sup}:
\begin{equation}
    C_{H} = -\beta \sum_{k=1}^{|\tilde{p}|} \mathcal{H}(p(\alpha |
    \tilde{p}_{<k})),
\end{equation}
where $\beta$ is a regularization coefficient, and $\alpha$ is a random variable
corresponding to a set of actions including all the outgoing edges and the stop
action. 

\subsection{Training Strategies: Reinforcement Learning}
The very same agent, NeuAgent, can be trained instead to maximize the final
reward without any supervision at each time step on which outgoing edge it
should follow. In this case, we only use a query $q$ from each example (see
Equation~\eqref{eq:example1}.)

Given a query, the NeuAgent navigates the graph, starting from the
starting node $s_S$, until it issues the \textit{stop} action or the number of hops
reaches the predefined maximum number of hops $N_h$. The last node $s_E$ in
which the agent was halted (either due to its own action or not) is considered
an answer node, and the reward $R(s_E, q)$ is computed. 

In this setup, we train the NeuAgent using the Q-Learning \cite{watkins1992q}
algorithm. The goal of Q-Learning is to learn to predict the expected reward for a given state-action pair, $Q(s_t,a)$, where $s_t$ is the current state the agent is in, and $a$ is one of the possible actions at the current state.

The optimal action-value function obeys the \textit{Bellman equation}, which
states that, given that the optimal value $Q^*(s_{t+1},a')$ of the sequences
$s'$ at the next time-step is known for all possible actions $a'$, the optimal
strategy can be obtained by selecting the action $a'$ of the next state
$s_{t+1}$ that maximizes the expected future reward $r_t + \gamma Q^*(s_{t+1},a')$,
\begin{equation}
    Q^*(s_t,a) = \Exp[r_t + \gamma \max_{a'}Q(s_{t+1},a') |s_t,a],
\end{equation}
where $\gamma$ is a discount factor for future rewards.

As in many reinforcement learning algorithms, the optimal policy can be achieved through the following update rule, derived from the Bellman equation:
\begin{equation}
    Q(s_t, a) \leftarrow Q(s_t, a) + \eta_t
            \left( r_t + \gamma \max_{a'} Q(s_{t+1},a') - Q(s_t, a) \right),
\end{equation}
where $\eta_t$ is the learning rate and $Q(s_{t+1},a')$ is known as the target Q-value.

The action-value $Q(s_t,a)$ can be estimated for each state-action pair
separately, which is impractical for our task due to the large state and action spaces, or it can be estimated using a function approximator. There are many choices for a function approximator and, in this work, we use the same neural network of the NeuAgent to do so, except that, instead of computing the probabilities for the actions as in the supervised case, the agent now computes the expected future rewards for the state-action pairs. In practice, the NeuAgent can be modified to output these expectations by simply replacing the exponential functions in eqs.~\eqref{eq:edge_p} and \eqref{eq:stop_p} by {\em sigmoid} functions, thus bounding the Q-values between zero and one, which are the reward limits:
\begin{equation}
Q(s_t,a) = Q(h_t,e_{t,j}) = \sigma \left(\phi_c(s_j)^\top \vh_t\right)
\end{equation}
\vspace{-8mm}
\begin{equation}
Q(s_t,\emptyset) = Q(h_t,\emptyset) = \sigma \left( \vv_{\emptyset}^\top \vh_t \right)
\end{equation}

The parameters $\theta$ can be learned by minimizing the
following loss function:
\begin{equation}
C_\text{RL} = \Exp_{\hat{s},\hat{a}}[(y - Q(\hat{s},\hat{a};\theta))^2],
\end{equation}
where $y=\Exp_{\hat{s},\hat{a}}[r + \gamma \max_{a'}Q(s_{t+1},a')]$.
Since $C_\text{RL}$ is differentiable with respect to $\theta$, this cost can be
minimized using SGD:
\begin{equation}
    \theta \leftarrow \theta - \eta \nabla C_{\text{RL}},
\end{equation}
where the gradients $\nabla C_\text{RL}$ are computed using backpropagation.

While $Q(s_t,a)$ is iteratively updated, the agent chooses the action with highest $Q(s_t,a)$ to maximize its expected future rewards. However, greedily following only the actions that promise maximum reward does not allow the agent to visit states whose initial predictions were associated with low rewards, but that could lead to higher rewards if visited. Thus, in order to balance the trade-off between exploration and exploitation, an $\epsilon$-greedy policy~\cite{sutton1998introduction} is used, in which a random action is selected at each step with probability $\epsilon$.

Following~\citet{mnih2015human}, we make use of the experience replay mechanism~\cite{lin1993reinforcement}, in which the agent's experiences are stored at each time step to be later sampled and used in the Q-learning updates. This smooths the training distribution and therefore lead to a more stable training and faster convergence. Prioritized Sweeping~\cite{moore1993prioritized} is used to sample the experiences, making experiences associated with positive rewards have a higher chance of being replayed than the ones associated with no reward. The technique makes convergence faster as the agent sees good examples more often.

\paragraph{Supervised + Reinforcement Learning:}
During the initial phase of training, most of the sampled traces will lead to a zero reward.\footnote{
    Note that the baseline is approximated using the Monte Carlo method, and during this initial phase, is likely to be zero.
} This problem is significantly amplified in the large-scale goal-driven web navigation where both of the state and action spaces are very large.
In order to avoid this issue of slow start, we first train an agent with
supervised learning, which helps put high probabilities correct/supervised
paths. Then, the agent is further fine-tuned with the Q-Learning algorithm, which teaches the agent how to deal with unseen nodes. Since the agent pretrained with supervised learning already learned reasonable policies, the exploration factor (i.e., $\epsilon$ in the $\epsilon$-greedy policy) can be set to a small fixed value throughout the Q-learning finetuning phase ($\epsilon=0.1$, in our experiments).

\section{Human Evaluation}

One unique aspect of the navigation task is that it is very difficult for
an average person who was not trained specifically for finding information by navigating through an information network.
There are a number of reasons behind this difficulty.
First, the person must be familiar with, via training, the graph structure of the network, and this often requires many months, if not years, of training.
Second, the person must have in-depth knowledge of a broad range of
topics in order to make a connection via different concepts between the themes and topics of a query to a target node.
Third, each trial requires the person carefully to read the whole content of the nodes as she navigates, which is a time-consuming and exhausting job.

Thus, unlike many other tasks in which the average human is often the upper-bound of the performance, the navigation task is challenging as well as interesting, as the progress in developing algorithms and models for artificial agents is not bounded by human intelligence. Nevertheless, in this work, we present the performance of human volunteers to put the performances of the proposed NeuAgents in perspective. 

We asked five volunteers to try up to 20 four-sentence-long queries\footnote{
    In a preliminary study with other volunteers, we found that, when the
    queries were shorter than $4$, they were not able to solve enough trials for
    us to have meaningful statistics.
    }
randomly selected from the test sets of \mbox{WikiNav-$\left\{ 4, 8, 16 \right\}$-4}
datasets. They were given up to two hours, and they were allowed to choose up to
the same maximum number of explored edges per node $N_n$ as the NeuAgents
(that is, $N_n=4$), and also were given the option to give up. The average reward
was computed as the fraction of correct trials over all the queries presented.

\section{Experiments and Analysis}
\label{sec:result}

\subsection{WikiNav: Quantitative Analysis}
\label{sec:wikiquant}

We report in Table~\ref{tab:webnav_results} the performance of the
NeuAgent-FF and NeuAgent-Rec models on the test set of all nine
\mbox{WikiNav-$\left\{ 4, 8, 16\right\}$-$\left\{ 1, 2, 4\right\}$} datasets. In addition to the proposed NeuAgents, we also report the results of the human evaluation.

We clearly observe that the level of difficulty is indeed negatively correlated with the query length $N_q$ but is positively correlated with the maximum number of allowed hops $N_h$.
The latter may be considered trivial, as the size of the search space grows exponentially with respect to $N_h$, but the former is not.
The former negative correlation confirms that it is indeed easier to solve the task with more information in a query. 

The NeuAgent-FF and NeuAgent-Rec shares similar performance when the maximum
number of allowed hops is small ($N_h=4$), but NeuAgent-Rec ((a) vs. (b))
performs consistently better for higher $N_h$, which indicates that having
access to history helps in long-term planning tasks. We also observe that the
larger and deeper NeuAgent-Rec ((b) vs. (c)) significantly outperforms the
smaller one, when a target node is further away from the starting node $s_S$.

Training with supervised learning and then finetuning the with
reinforcement learning improves the performance even further ((e) vs. (f)).
Contrary to \citet{mnih2015human}, who had to
freeze the weights of the target network $Q(s_{t+1},a')$ for many minibatch
updates to achieve stability, using the parameters of the previous iteration for the target network was enough to achieve a stable training in our experiments.
We conjecture that this is because the weights of the pretrained network are already at a good local minimum, making training more stable and thus eliminating the need for the trick.

The performance difference between simpler and more sophisticated models is
more evident as the difficulty of the task increases ($N_q \downarrow$ and $N_h \uparrow$). For instance, the best performing models in (e) and (f) use a multi-layer LSTM and attention-based representation for query and content.

\begin{figure}
    \centering
    \includegraphics[width=0.99\columnwidth]{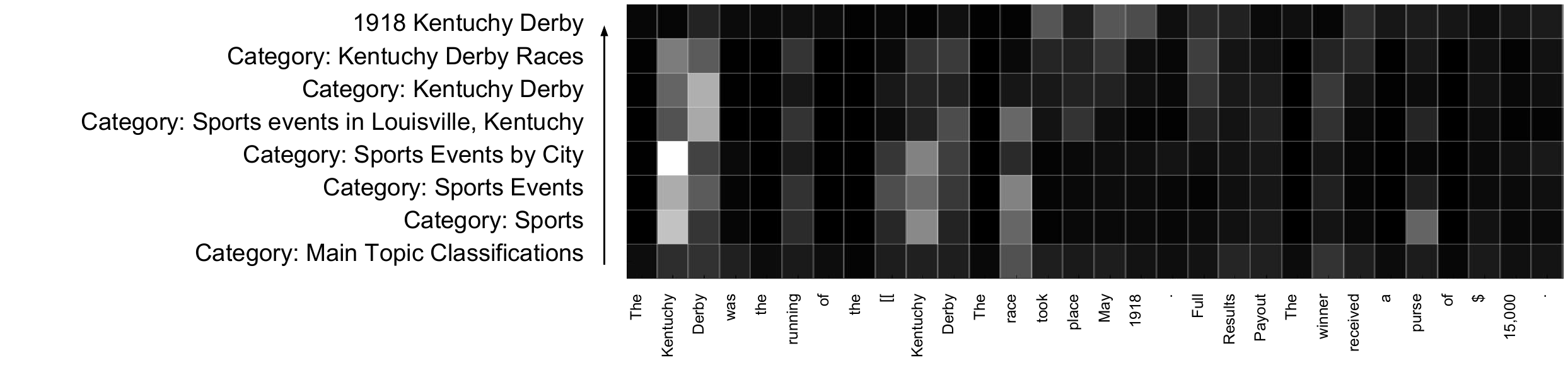}
    \caption{Visualization of the attention mechanism over a sample query. The
    horizontal axis corresponds the words in the input query, the vertical axis
    corresponds to the title of the current Wikipedia article, and the brighter
    the cell, the higher the attention weight.}
    \label{fig:query_att_vis}
\end{figure}

In Figure~\ref{fig:query_att_vis}, we present an example of how the attention weights over the query words dynamically evolve as the model navigates toward a target node. We note that higher weights are assigned the to most relevant words in the query (in the example, Kentucky, Derby, and race) that will lead to the correct document.

The human participants generally performed worse than the NeuAgents. We
attribute this to a number of reasons. First, the NeuAgents are trained
specifically on the target domain (Wikipedia), while the human participants have not been. Second, we observed that the volunteers were rapidly exhausted from reading multiple articles in sequence.

\begin{table}
    \centering
    \begin{tabular}{c|cccccccc}
        Model & Training & $f_{\text{core}}$ & \#Layers $\times$ \#Units & $\phi_q$ & $\phi_c$ \\
        \toprule
        (a) & Sup & $f_{\text{ff}}$ & $1\times512$ $\tanh$ & BoW & BoW \\
        (b) & Sup & $f_{\text{rec}}$ & $1\times512$  LSTM & BoW & BoW \\
        (c) & Sup & $f_{\text{rec}}$ & $8\times2048$ LSTM & BoW &  BoW \\
        (d) & Sup & $f_{\text{rec}}$ & $8\times2048$
    LSTM & Att & BoW \\
        (e) & Sup & $f_{\text{rec}}$ & $8\times2048$
    LSTM & Att & Att \\
        (f) & Sup + RL & $f_{\text{rec}}$ & $8\times2048$
    LSTM & Att & Att \\
    \end{tabular}
    \newline
    \vspace*{0.5 cm}
    \newline
    \begin{tabular}{c|ccc|ccc|ccc}
        & \multicolumn{3}{c|}{$N_q=1$} & \multicolumn{3}{c|}{2} & \multicolumn{3}{c}{4} \\
        \hline
        Model & $N_h=4$ & 8 & 16 & 4 & 8 & 16 & 4 & 8 & 16 \\
        \toprule
        (a) & 21.5 & 4.7 & 1.2 & 40.0 & 9.2 & 1.9 & 45.1 & 12.9 & 2.9 \\
        (b) & 22.0 & 5.1 & 1.7 & 41.1 & 9.2 & 2.1 & 44.8 & 13.3 & 3.6 \\
        (c) & 17.7 & 10.9 & 8.0 & 35.8 & 19.9 & 13.9 & 39.5 & 28.1 & 21.9 \\
        (d) & 22.9 & 15.8 & 12.5 & 41.7 & 24.5 & 17.8 & 46.8 & 34.2 & 28.2 \\
        (e) & - & - & - & - & - & - & 47.3 & 35.0 & 29.9 \\
        (f) & - & - & - & - & - & -  & {\bf 49.8} & {\bf 36.1} & {\bf 30.9} \\
        Humans & - & - & - & - & - & - & 14.5 & 8.8 & 5.0 

    \end{tabular}
    \caption{The average reward by the NeuAgents and humans on the
        test sets of \mbox{WikiNav-$N_h$-$N_q$}.
    }
    \label{tab:webnav_results}
\end{table}

\subsection{WikiNav: Qualitative Analysis}
\label{sec:qualitative}

In Table~\ref{tab:qualitative}, we present a few example runs by NeuAgent
model (d) from Table~\ref{tab:webnav_results} trained on the WebNav-8-1.
In those two successful runs, we see that the agent was able to plan its trajectory correctly.
For instance, in the second successful example, the agent starts with a broader theme of the query sentence, ``government'' and narrows down toward more specific themes (i.e., ``the United States'' $\to$ ``the Confederate States'' $\to$ its ``electoral
college''). 

Even in the cases of failure, we observe that the agent is able to navigate
through relevant nodes rather than going completely off the topic. Again, the second failed run exhibits a pattern that is intuitively understandable. There are two major themes in the query sentence, which are ``random process'' and ``human mobility.''
The agent starts by the theme of ``random process,'' following through nodes related to ``applied mathematics.''
The random process in this query was described as ``predictable,'' and the agent correctly noticed that ``stable process'' which can be considered ``predictable.''
However, the agent failed to find a page in which the remaining theme (``human mobility'') occurs together with this ``predictable random process.'' 

These examples illustrate that to find correct nodes the agent must have at least partial understanding of how terms relate to each other, which promises a large potential for using this web navigation agent as a retrieval mechanism.

\begin{table}
    \footnotesize
    \centering
    \begin{tabular}{l |  l}
        \hline
        \hline
        {\bf Query} & Young adults are the most likely age group to smoke, with a
        marked decline \\
        & in smoking rates with increasing age. \\
        \hline
        {\bf Trace} & 
            category:health $\to$
            category:education by country $\to$
            category:smoking by country 
            \\
            &
            $\to$ list of countries by cigarette consumption per capita
        \\
        \hline
        \hline
        {\bf Query} & This system was established by the [[Confederate States
        Constitution]], in \\ 
        & emulation of the [[United States Constitution]]. \\
        \hline
        {\bf Trace} & 
            category:government $\to$
            category:government by country $\to$ \\
            & category:government of the confederate states of america $\to$ \\
            & electoral college (confederate states) \\
        \hline
        \hline
    \end{tabular}

    (a) Successful Runs
    \vspace{3mm}

    \begin{tabular}{l |  l}
        \hline
        \hline
        {\bf Query} & Other types of exercise include the TEWT (Tactical Exercise
        Without Troops), \\
        & also known as a [[sand table]], map or cloth model exercise \\
        \hline
        {\bf Trace} & 
            category:sports $\to$
            category:physical exercise $\to$
            category:bodyweight exercise $\to$ \\
            & range of motion (exercise machine) \\
        \hline
        \hline
        {\bf Query} & Predictability  Although the human mobility is modeled as
        a random process, it is \\
        & surprisingly predictable. \\
        \hline
        {\bf Trace} & 
        category:creativity $\to$
        category:applied mathematics $\to$ \\
        & category:mathematical finance $\to$
        stable process \\
        \hline
        \hline
    \end{tabular}

    (b) Failed Runs

    \caption{Traces generated by the NeuAgent-Rec trained on \mbox{WikiNav-8-1} using the queries from the test set. We present two examples per each of (a) successful and (b) failed runs.}
    \label{tab:qualitative}
\end{table}

\subsection{WikiNav-Jeopardy}
\label{sec:resultsjeo}

\paragraph{Settings:} 
We test the model (d) from Section~\ref{sec:wikiquant} (NeuAgent-Rec with eight layers of 2048 LSTM units and the attention-based query representation) on the WikiNav-Jeopardy.
We evaluate two training strategies.
The first strategy is straightforward supervised learning, in which we train a NeuAgent-Rec on WikiNav-Jeopardy from scratch.
In the other strategy, we pretrain a NeuAgent-Rec first on the WikiNav-16-4 and fine-tune it on WikiNav-Jeopardy.

We compare the proposed NeuAgent against three search strategies. The first one,
{\it SimpleSearch}, is a simple inverted index-based strategy. SimpleSearch
scores each Wikipedia article by the TF-IDF weighted sum of words that co-occur
in the articles and a query and returns top-$K$ articles. Second, we use Lucene,
a popular open source information retrieval library, in its default
configuration with BM25 ranking function on the whole Wikipedia corpus. Lastly, we use Google Search
API\footnote{
    \url{https://cse.google.com/cse/}
}, while restricting the domain to \url{wikipedia.org}.

Each system is evaluated by document recall at $K$ (Recall@$K$) and Mean Reciprocal Rank (MRR) since there is only one relevant document per question. We vary $K$ to be 1, 5, or 100. 
In the case of the NeuAgent, we run beam search with width set to $K$ and returns all the $K$ final nodes to compute the document recall. 

\begin{table}[htb]
    \centering
    \begin{tabular}{l c ||c|c|c|c}
        Model & Pre$^\star$ & MRR & Recall@1 & Recall@5 & Recall@100 \\ 
        \toprule
        NeuAgent & & 16.9 & 13.9 & 20.2 & 45.2 \\ 
        NeuAgent & $\checkmark$ & \textbf{20.3} & \textbf{18.9} & \textbf{24.1} & \textbf{47.6} \\ 
        \midrule
        \multicolumn{2}{l||}{SimpleSearch} & 7.8 & 5.4 & 12.6 & 42.1 \\ 
        \multicolumn{2}{l||}{Lucene (BM25)} & 9.3 & 6.3 & 14.7 & 46.6 \\ 
        \multicolumn{2}{l||}{Google} & 17.6 & 14.0 & 22.1 & 39.8 \\ 
    \end{tabular}

    \caption{Results on WikiNav-Jeopardy. In bold we show statistically significant results ($p < 0.05$) according to Student's paired t-test with a Bonferroni correction (code modified from \url{https://github.com/castorini/Anserini/blob/master/src/main/python/compare_runs.py}). $(\star)$ Pretrained on WikiNav-16-4.}
    \label{tab:jeoresults}
\end{table}

\paragraph{Result and Analysis:} In Table~\ref{tab:jeoresults}, we report the results on WikiNav-Jeopardy. The proposed NeuAgent clearly outperforms all the three search-based strategies,
when it was pretrained on the WikiNav-16-4. The superiority of the pretrained
NeuAgent is more apparent when the number of candidate documents is constrained
to be small, implying that the NeuAgent is able to rank a correct target article accurately.  Although the NeuAgent performs comparably to the other search-based strategy even without pretraining, the benefit of pretraining on the much larger WikiNav is clear.

We emphasize that these search-based strategies have access to all the nodes for each input query. The NeuAgent, on the other hand, only observes the nodes as it visits during navigation. This success clearly demonstrates a potential in using the proposed NeuAgent pretrained with a
dataset compiled by the proposed WebNav for the task of focused crawling~\cite{chakrabarti1999focused,alvarez2007deepbot}.

\section{Related Work}
\label{sec:related}

\subsection{Goal-Driven Web Navigation}

This work is indeed not the first to notice the possibility of a website, or
possibly the whole web, as a world in which intelligent agents, including
ourselves, explore to achieve a certain goal. One most relevant recent work to
ours is perhaps Wikispeedia from
\citet{west2009wikispeedia,west2012automatic,west2012human}.

West~et~al. proposed the following game, called Wikispeedia. The game's world is nearly identical to the goal-driven navigation task proposed in this work. More specifically, they converted ``Wikipedia for Schools'',\footnote{
    \url{http://schools-wikipedia.org/}
}, which contains approximately 4,000 articles and 120,000 hyperlinks as of
2008, into a graph whose nodes are articles and directed edges are hyperlinks.
From this graph, a pair of nodes is randomly selected and provided to an agent, be it a person or an artificial agent. 

The agent's goal is to start from the first node, navigate the graph and reach the second (target) node. Similarly to the WikiNav, the agent has access to the text content of the current nodes and all the immediate neighboring nodes. One major difference is that the target is given as a whole page rather than a sentence, meaning that there is a single target node in the Wikispeedia while there may be multiple target nodes in the proposed WikiNav or any goal-driven web navigation created by the WebNav. 

From this description, we see that the goal-driven web navigation is a
generalization and re-framing of the Wikispeedia by West~et~al. First, we let a query contain less information, making it much more difficult for an agent to navigate to a target node without language understanding and planning capabilities. Furthermore, a major research question by
\citet{west2012human} was to ``{\it 
    understand how humans navigate and find the information they are looking for
},'' whereas in this work we are focused on building novel and better retrieval mechanisms.

Recently \citet{narasimhan2015language} proposed to
incorporate natural language understanding and planning into a single problem.
They consider multi-user dungeon (MUD) games as a target task, in which the
world is only partially observed as natural language instructions.  Furthermore,
the actions are often defined by natural language sentences as well.

The proposed goal-driven web navigation, more specifically WikiNav, is similar
to MUD games. A major difference is in the complexity of the task. For instance,
the goal-driven web navigation built from a real website, such as the WikiNav
proposed here, uses a vocabulary of approximately 370k unique words, while
that of the ``Fantasy World'' from \citet{narasimhan2015language} contains
a substantially smaller number of words (1,340). 

Also related are the personalized PageRank algorithms~\cite{kamvar2003exploiting,jeh2003scaling,haveliwala2002topic,haveliwala2003analytical}. The idea is to bias PageRank vectors according to user profiles, topics, or queries. The large cost of computing these vectors poses a challenge in tasks where the bias vector frequently changes, such as in each new question of a question-answering task. Our agent scales well in an online setting as only a few dozen links are evaluated at each navigational step.

\subsection{Focused Crawling}

Agents trained on the task of large-scale goal-driven web navigation can be readily applied to a number of applications. Perhaps the most important one is to use any technology built for solving this task as a focused crawler, or part of it.
A focused crawler aims at crawling websites with a predefined, specific topic, unlike traditional crawlers whose aim is to index all possible web pages \cite{chakrabarti1999focused,alvarez2007deepbot} that later will be retrieved based on the terms present in the input query and certain metrics, such as PageRank \cite{page1999pagerank}.
This is an interesting problem, as much of the content available on the Internet is either hidden or dynamically generated, meaning that they need to be searched on-the-fly \cite{alvarez2007deepbot}.
Focused crawling is also more efficient since fewer documents are visited and analyzed.
If we consider the query in the proposed goal-driven web navigation as an unstructured form of topics, the agent trained to solve the goal-driven navigation can readily be applied to this focused crawling.
Our method can, additionally, learn textual representations and crawling strategies directly from data in an end-to-end training process.

The idea of learning to navigate information networks conditioned on a query was proposed in various earlier works.
\citet{rennie1999using}, for example, introduced an agent that can learn the focused crawling task through reinforcement learning. However, their method was evaluated in two small datasets restricted to the domains of computer science departments and company websites, and the queries were single topics.
We, instead, approach the focused crawling task by making our agent navigate a much larger informational graph and use complex natural language questions from the Jeopardy! game as queries.

Along the same lines, \citet{meusel2014focused} use online-based classification algorithms in combination with a bandit-based selection strategy to efficiently crawl pages with markup languages such as RDFa, Microformats, and Microdata. In contrast, our agent makes its decisions based only on the natural language description of the web pages, without the need for structured content.

\section{Summary}
\label{sec:summary}

In this chapter, we described a large-scale goal-driven web navigation agent and argue that it serves as an alternative for index-based search engines. We release
a software tool, called WebNav, that compiles a given website into a goal-driven
web navigation task. As an example, we construct WikiNav from Wikipedia using
WebNav. We extend WikiNav with {\it Jeopardy!} questions, thus creating
WikiNav-Jeopardy.
We evaluate various neural net based agents on WikiNav, an information retrieval task, and WikiNav-Jeopardy, a question-answering task.
Our results show that our agent pretrained on WikiNav outperforms two strong inverted index-based search engines on the WikiNav-Jeopardy. These empirical results support our claim on the usefulness of the navigation agents in challenging applications such as document retrieval, question answering, and focused crawling.



\Chapter{Retrieving from a Black Box}{Query Reformulation with Reinforcement Learning}
\label{chap:query_reformulation}

In the previous chapter, we described a retrieval agent that navigates a web of documents to find information. Despite its novelty, this agent requires a corpus of densely-linked documents to work well. Wikipedia has this property but the web, in general, lack of. Training an agent to handle all sorts of different web-page designs, especially dynamic-generated content, would require an enormous engineering effort. What if, instead, we could use search engines to do the heavy lifting of crawling, storing, and retrieving information in a standard format that a learning agent would then ingest and produce new results? This is possible, but there are some problems with existing search engines that we must be aware of. For example, when we request some information using a long or inexact description of it, these systems often fail to deliver relevant items. In this case, what typically follows is an iterative process in which we try to express our need differently in the hope that the system will return what we want. This is a major issue in information retrieval. For instance,~\citet{huang2009analyzing} estimate that 28-52\% of all the web queries are modifications of previous ones.

\begin{figure}[ht]
\begin{center}
\centerline{\includegraphics[width=0.8\columnwidth]{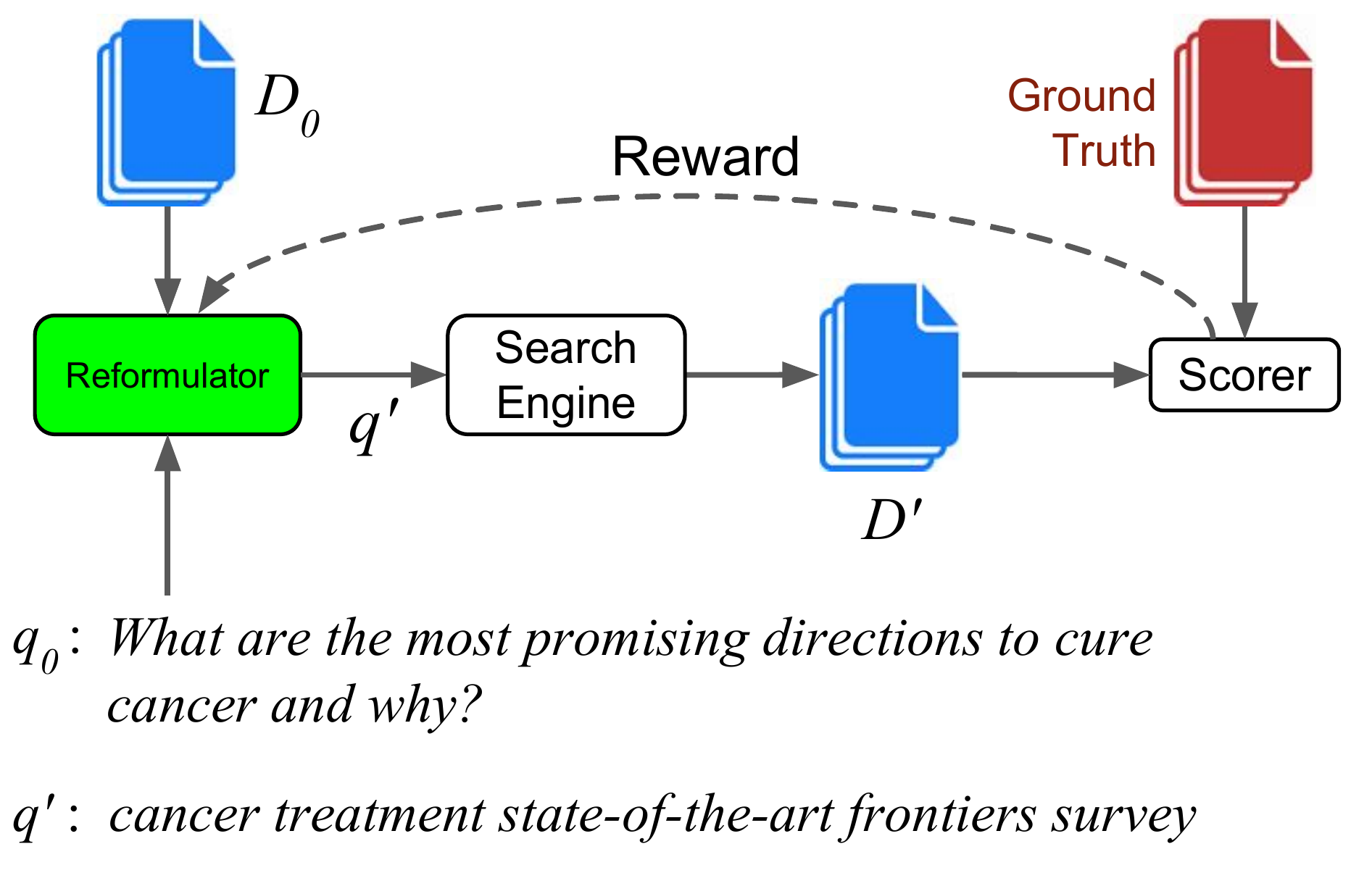}}
\caption{A graphical illustration of the proposed framework for query reformulation. A set of documents $D_0$ is retrieved from a search engine using the initial query $q_0$. Our reformulator selects terms from $q_0$ and $D_0$ to produce a reformulated query $q'$ which is then sent to the search engine. Documents $D'$ are returned, and a reward is computed against the set of relevant documents.
The reformulator is trained with reinforcement learning to produce a query, or a series of queries, to maximize the expected return.
}
\label{fig:query_reformulator}
\end{center}
\end{figure}

To a certain extent, this problem occurs because search engines rely on matching terms in the query with terms in documents, to perform retrieval. If there is a mismatch between them, a relevant document may be missed.


One way to address this problem is to automatically rewrite a query so that it becomes more likely to retrieve relevant documents. This technique is known as \textit{automatic query reformulation}. It typically expands the original query by adding terms from, for instance, dictionaries of synonyms such as WordNet~\cite{miller1995wordnet}, or from the initial set of retrieved documents~\cite{xu1996query}. This latter type of reformulation is known as pseudo (or blind) relevance feedback (PRF), in which the relevance of each term of the retrieved documents is automatically inferred.

The proposed method is built on top of PRF but differs from previous works as we frame the query reformulation problem as a reinforcement learning (RL) problem. An initial query is the natural language expression of the desired goal, and an agent (i.e., reformulator) \textit{learns} to reformulate an initial query to maximize the expected return (i.e., retrieval effectiveness) through actions (i.e., selecting terms for a new query). The environment is a search engine which produces a new state (i.e., retrieved documents). Our framework is illustrated in Figure~\ref{fig:query_reformulator}.

The most important implication of this framework is that a search engine is treated as a \textit{black box} that an agent learns to use in order to retrieve more relevant items. This opens the possibility of training an agent to use a search engine for a task other than the one it was originally intended for. To support this claim, we evaluate our agent on the task of question answering (Q\&A), citation recommendation, and passage/snippet retrieval.

As for training data, we use two publicly available datasets (TREC-CAR and Jeopardy) and introduce a new one (MS Academic) with hundreds of thousands of \textit{query}/\textit{relevant document} pairs from the academic domain.

Furthermore, we present a method to estimate the upper bound effectiveness of our RL-based model. Based on the estimated upper bound, we claim that this framework has a strong potential for future improvements.

Here we summarize our main contributions:
\begin{itemize}
\setlength\itemsep{1pt}
\item A reinforcement learning framework for automatic query reformulation.
\item A simple method to estimate the upper-bound effectiveness of an RL-based model in a given environment.
\item A new large dataset with hundreds of thousands of \textit{query}/\textit{relevant document} pairs.\footnote{The dataset and code to run the experiments are available at \url{https://github.com/nyu-dl/QueryReformulator}.}
\end{itemize}

\begin{figure*}[ht]
\begin{center}
\centerline{\includegraphics[width=0.7\columnwidth]{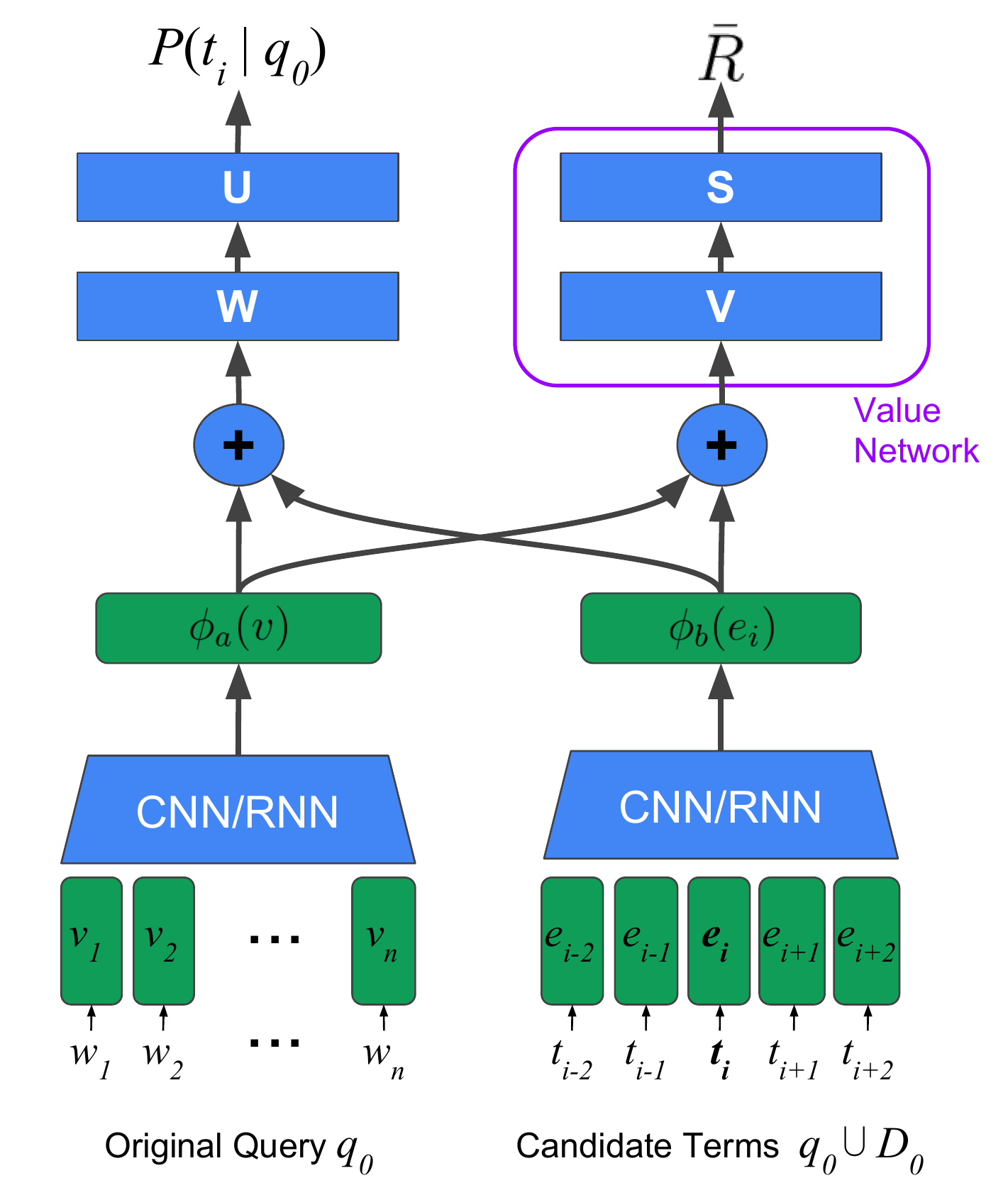}}
\caption{An illustration of our neural network-based reformulator.}
\label{fig:reformulator}
\end{center}

\end{figure*} 

\section{A Reinforcement Learning Approach}

\subsection{Model Description}
\label{sec:rl}

In this section, we describe the proposed method, illustrated in Figure~\ref{fig:reformulator}.

The inputs are a query $q_0$ consisting of a sequence of words
$(w_1, ..., w_n)$ and a candidate term $t_i$ with some context words $(t_{i-k},\allowbreak ...,\allowbreak t_{i+k})$, where $k \geq 0$ is the context window size. Candidate terms are from $q_0 \cup D_0$, the union of the terms in the original query and those from the documents $D_0$ retrieved using $q_0$.

We use a dictionary of pretrained word embeddings~\cite{mikolov2013efficient} to convert the symbolic terms ${w_j}$ and ${t_i}$ to their vector representations $v_j$ and $e_i \in \RR^d$, respectively. We map out-of-vocabulary terms to an additional vector that is learned during training.

We convert the sequence $\{v_j\}$ to a fixed-size vector $\phi_a(v)$ by using either a Convolutional Neural Network (CNN) followed by a max pooling operation over the entire sequence~\cite{kim2014convolutional} or by using the last hidden state of a Recurrent Neural Network (RNN).\footnote{To deal with variable-length inputs in a mini-batch, we pad smaller ones with zeros on both ends so they end up as long as the largest sample in the mini-batch.}

Similarly, we fed the candidate term vectors ${e_i}$ to a CNN or RNN to obtain a vector representation ${\phi_b(e_i)}$ for each term $t_i$. The convolutional/recurrent layers serve an important role in capturing context information, especially for out-of-vocabulary and rare terms. CNNs can process candidate terms in parallel, and, therefore, are faster for our application than RNNs. RNNs, on the other hand, can encode longer contexts.

Finally, we compute the probability of selecting $t_i$ as:
\begin{equation} \label{eq:1}
P(t_i|q_0) = \sigma( U^\mathsf{T} \tanh( W(\phi_a(v) \Vert \phi_b(e_i)) + b )),
\end{equation}
where $\sigma$ is the sigmoid function, $\Vert$ is the vector concatenation operation, $W \in \RR^{d \times 2d}$ and $U \in \RR^{d}$ are weights, and $b \in \RR$ is a bias.

At test time, we define the set of terms used in the reformulated query as $T=\{t_i\ |\ P(t_i|q_0)>\epsilon\}$, where $\epsilon$ is a hyperparameter.
At training time, we sample the terms according to their probability distribution:
\begin{equation}
T=\{t_i\ |\ \alpha=1 \wedge \alpha \sim P(t_i|q_0)\}.
\end{equation}
We concatenate the terms in $T$ to form a reformulated query $q'$, which will then be used to retrieve a new set of documents $D'$. 

\subsection{Sequence Generation}
\label{sec:seqgen}
One problem with the method previously described is that terms are selected independently. This may result in a reformulated query that contains duplicated terms since the same term can appear multiple times in the feedback documents. Another problem is that the reformulated query can be very long, resulting in a slow retrieval.

To solve these problems, we extend the model to sequentially generate a reformulated query, as proposed by~\citet{aqa-iclr:2018}. We use a Recurrent Neural Network (RNN) that selects one term at a time from the pool of candidate terms and stops when a special token is selected. The advantage of this approach is that the model can remember the terms previously selected through its hidden state. It can, therefore, produce more concise queries.

We define the probability of selecting $t_i$ as the k-th term of a reformulated query as:
\begin{equation}
P(t_i^k|q_0) \propto \exp(\phi_b(e_i)^\mathsf{T} h_k),
\end{equation}
where $h_k$ is the hidden state vector at the k-th step, computed as:
\begin{equation}
h_k = \tanh(W_a \phi_a(v) + W_b \phi_b(t^{k-1}) + W_h h_{k-1}),
\end{equation}
where $t^{k-1}$ is the term selected in the previous step and $W_a \in \RR^{d \times d}$, $W_b \in \RR^{d \times d}$, and $W_h \in \RR^{d \times d}$ are weight matrices. In practice, we use an LSTM~\cite{hochreiter1997long} to encode the hidden state as this variant is known to perform better than a vanilla RNN.

We avoid normalizing over a large vocabulary by using only terms from the retrieved documents. This makes inference faster and training practical since learning to select words from the whole vocabulary might be too slow with reinforcement learning, although we leave this experiment for the future.

\subsection{Training}

We train the proposed model using REINFORCE~\cite{williams1992simple} algorithm. The per-example stochastic objective is defined as
\begin{equation} \label{eq:2}
    C_a = (R - \bar{R}) \sum_{t \in T} -\log P(t | q_0),
\end{equation}
where $R$ is the reward, and $\bar{R}$ is the baseline,
computed by the value network as:
\begin{equation} \label{eq:3}
	\bar{R} = \sigma( S^\mathsf{T} \tanh(V ( \phi_a(v) \Vert \bar{e} ) + b)),
\end{equation}
where $\bar{e} = \frac{1}{N} \sum_{i=1}^{N} \phi_b(e_i)$, $N=|q_0 \cup D_0|$, $V \in \RR^{d \times 2d}$ and $S \in \RR^{d}$ are weights and $b \in \RR$ is a bias.
We train the value network to minimize
\begin{equation} \label{eq:4}
	C_b = \alpha||R-\bar{R}||^2,
\end{equation}
where $\alpha$ is a small constant (e.g., 0.1) multiplied to the loss in order to stabilize learning. We conjecture that the stability is due to the slowly evolving value network, which directly affects the learning of the policy. This effectively prevents the value network from fitting extreme cases (unexpectedly high or low reward.)

We minimize $C_a$ and $C_b$ using stochastic gradient descent (SGD) with the gradient computed by backpropagation~\cite{rumelhart1988learning}. This allows the entire model to be trained end-to-end directly to optimize the retrieval effectiveness. 

\subparagraph{Entropy Regularization:} Similar to experiments in Chapter~\ref{chap:webnav}, we observed that the probability distribution in Equation~\eqref{eq:1} became highly peaked in preliminary experiments. This phenomenon led to the trained model not being able to explore new terms that could lead to a better-reformulated query. We address this issue by regularizing the negative entropy of the probability distribution. We add the following regularization term to the original cost function in Equation~\eqref{eq:2}:
\begin{equation}
	C_H = -\lambda \sum_{t \in q_0 \cup D_0} P(t|q_0)\log P(t|q_0),
\end{equation}
where $\lambda$ is a regularization coefficient.

\section{Related Work}

Query reformulation techniques are either based on a global method, which ignores a set of documents returned by the original query, or a local method, which adjusts a query relative to the documents that initially appear to match the query. In this work, we focus on local methods.

A popular instance of a local method is the \textit{relevance model}, which incorporates pseudo-relevance feedback into a language model form~\cite{lavrenko2001relevance}. The probability of adding a term to an expanded query is proportional to its probability of being generated by the language models obtained from the original query and from the document in which the term occurs. This framework has the advantage of not requiring pairs of query and relevant document as training data since inference is based on word co-occurrence statistics.

Unlike the relevance model, algorithms can be trained with supervised learning, as proposed by~\citet{cao2008selecting}. A training dataset is automatically created by labeling each candidate term as relevant or not based on their individual contribution to the retrieval effectiveness. Then a binary classifier is trained to select expansion terms. In Section~\ref{sec:qr_experiments}, we present a neural network-based implementation of this supervised approach.

An alternative for this supervised framework is to \textit{iteratively} reformulate the query by selecting one candidate term at each retrieval step. This can be viewed as navigating a graph where the nodes represent queries and associated retrieved results and edges exist between nodes whose queries are simple reformulations of each other~\cite{diaz2016pseudo}. However, it can be slow to reformulate a query this way as the search engine must be queried for each newly added term. Our method, on the contrary, queries the search engine with various new terms at once.

Another technique based on supervised learning is to learn a common latent representation of queries and relevant documents terms by using a \textit{click-through} dataset~\cite{sordoni2014learning}. Neighboring document terms of a query in the latent space are selected to form an expanded query. Instead of using a \textit{click-through} dataset, which is often proprietary, it is possible to use an alternative dataset consisting of pairs of web page title and anchor text. 

Queries can also be expanded with terms that are close in the embedding space~\cite{roy2016using,kuzi2016query}. In this case, word embeddings trained with feedback documents as negative examples result in expanded queries that are more effective than when embeddings trained with corpus-level negatives are used~\cite{diaz2016query}. 

Perhaps the closest work to ours is that by \citet{narasimhan2016improving}, in which a reinforcement learning based approach is used to reformulate queries iteratively.
A key difference is that in their work, the reformulation component uses domain-specific template queries. Our method, on the other hand, assumes open-domain queries. 


\section{Experiments}
\label{sec:qr_experiments}
In this section, we describe our experimental setup, including baselines against which we compare the proposed method, metrics, reward for RL-based models, datasets, and implementation details.

\subsection{Baseline Methods}

\subparagraph{Raw:} The original query is given to a search engine without any modification. We evaluate two search engines in their default configuration: Lucene\footnote{https://lucene.apache.org/} with BM25~\cite{robertson1995okapi}  as the ranking function (Raw-BM25) and Google Search\footnote{https://cse.google.com/cse/} (Raw-Google). 

\subparagraph{Pseudo-Relevance Feedback (PRF-TFIDF):}
A query is expanded with terms from the documents retrieved by a search engine using the original query. In this work, the top-$N$ TF-IDF terms from each of the top-$K$ retrieved documents are added to the original query, where $N$ and $K$ are selected by a grid search on the validation data.

\subparagraph{PRF-Relevance Model (PRF-RM):} This is our implementation of the relevance model for query expansion~\cite{lavrenko2001relevance}. The probability of adding a term $t$ to the original query is given by:
\begin{equation}
P(t|q_0) = (1-\lambda) P'(t|q_0) + \lambda \sum_{d \in D_0} P(d) P(t|d) P(q_0|d),
\end{equation}
where $P(d)$ is the probability of retrieving the document $d$, assumed uniform over the set, $P(t|d)$ and $P(q_0|d)$ are the probabilities assigned by the language model obtained from $d$ to $t$ and $q_0$, respectively. $P'(t|q_0)= \frac{\text{tf}(t \in q)}{|q|}$, where $\text{tf}(t,d)$ is the term frequency of $t$ in $d$. We set the interpolation parameter $\lambda$ to 0.65, which was the best value found by a grid-search on the development set.

We use a Dirichlet smoothed language model~\cite{zhai2001study} to compute a language model from a document $d \in D_0$:
\begin{equation}
P(t|d)=\frac{\text{tf}(t,d)+u P(t|C)}{|d| + u},
\end{equation}
where $u$ is a scalar constant ($u=1500$ in our experiments), and $P(t|C)$ is the probability of $t$ occurring in the entire corpus $C$.

We use the $N$ terms with the highest $P(t|q_0)$ in an expanded query, where $N=100$ was the best value found by a grid-search on the development set. 

\subparagraph{Embeddings Similarity:}
Inspired by the methods proposed by~\citet{roy2016using} and~\citet{,kuzi2016query}, the top-$N$ terms are selected based on the cosine similarity of their embeddings against the original query embedding. 
Candidate terms come from documents retrieved using the original query (PRF-Emb), or from a fixed vocabulary (Vocab-Emb). We use pretrained embeddings from~\citet{mikolov2013efficient}, and it contains 374,000 words.

\subsection{Proposed Methods}
\label{sec:proposed_methods}

\subparagraph{Supervised Learning (SL):} Here we detail a deep learning-based variant of the method proposed by~\citet{cao2008selecting}. It assumes that query terms contribute independently to the retrieval effectiveness. We thus train a binary classifier to select a term if the retrieval effectiveness increases beyond a preset threshold when that term is added to the original query. More specifically, we mark a term as relevant if $(R' - R) / R > 0.005$, where $R$ and $R'$ are the retrieval effectiveness of the original query and the query expanded with the term, respectively.

We experiment with two variants of this method: one in which we use a convolutional network for both original query and candidate terms (SL-CNN), and the other in which we replace the convolutional network with a single hidden layer feed-forward neural network (SL-FF). In this variant, we average the output vectors of the neural network to obtain a fixed size representation of $q_0$.

\subparagraph{Reinforcement Learning (RL):} We use multiple variants of the proposed RL method. RL-CNN and RL-RNN are the models described in Section~\ref{sec:rl}, in which the former uses CNNs to encode query and term features, and the latter uses RNNs (more specifically, bidirectional LSTMs). RL-FF is the model in which term and query vectors are encoded by a single hidden layer feed-forward neural network. In the RL-RNN-SEQ model, we add the sequential generator described in Section~\ref{sec:seqgen} to the RL-RNN variant.

\begin{table*}

\begin{center}
\begin{scriptsize}
\begin{tabular}{lcc|ccc|cc|cc}
 &  &  & \multicolumn{3}{c|}{Queries} & \multicolumn{2}{c|}{Relevant Docs/Query} & \multicolumn{2}{c}{Words/Doc}\\
Dataset &Corpus &Docs & Train & Valid & Test & Avg. & Std. & Avg. & Std.\\
\noalign{\vskip 1mm}
\hline
\noalign{\vskip 1mm}
TREC-CAR & Wikipedia Paragraphs & 3.5M & 585k & 195k & 195k & 3.6 & 5.7 & 84 & 68\\
Jeopardy & Wikipedia Articles & 5.9M & 118K & 10k & 10k & 1.0 & 0.0 & 462 & 990\\
MSA & Academic Papers & 480k & 270k & 20k & 20k & 17.9 & 21.5 & 165 & 158\\
\end{tabular}
\end{scriptsize}
\end{center}
\vskip -2mm
\caption{Summary of the datasets.}
\label{tab:datasets}

\end{table*}

\subsection{Datasets}

We summarize in Table~\ref{tab:datasets} the datasets. 

\subparagraph{TREC - Complex Answer Retrieval (TREC-CAR):} This is a publicly available dataset automatically created from Wikipedia whose goal is to encourage the development of methods that respond to more complex queries with longer answers~\cite{dietz2017trec}. A query is the concatenation of an article title and one of its section titles. The relevant documents are the paragraphs within that section. For example, a query is ``\textit{Sea Turtle, Diet}'' and the relevant documents are the paragraphs in the section ``\textit{Diet}'' of the ``\textit{Sea Turtle}'' article. The corpus consists of all the English Wikipedia paragraphs, except the abstracts. The released dataset has five predefined folds, and we use the first three as the training set and the remaining two as validation and test sets, respectively.

\subparagraph{Jeopardy:} This is a publicly available Q\&A dataset introduced in Chapter~\ref{chap:webnav}. A query is a question from the \textit{Jeopardy!} TV Show, and the corresponding document is a Wikipedia article whose title is the answer. For example, a query is \textit{``For the last eight years of his life, Galileo was under house arrest for espousing this man’s theory''} and the answer is the Wikipedia article titled \textit{``Nicolaus Copernicus''}.
The corpus consists of all the articles in the English Wikipedia.

\subparagraph{Microsoft Academic (MSA):} This dataset consists of academic papers crawled from Microsoft Academic API.\footnote{https://www.microsoft.com/cognitive-services/en-us/academic-knowledge-api} The crawler started at the paper~\citet{silver2016mastering} and traversed the graph of references until 500,000 papers were crawled. We then removed papers that had no reference within or whose abstract had less than 100 characters. We ended up with 480,000 papers.

A query is the title of a paper, and the relevant answer consists of the papers cited within. Each document in the corpus consists of its title and abstract.\footnote{This was done to avoid the large computational cost for indexing and searching full papers.}
This dataset differs from the other two in that it uses different corpus and terminologies, and it has more relevant documents per query, thus favoring reformulation methods that produce more comprehensive queries.

\begin{table*}[t]
\begin{center}
\begin{small}
\begin{tabular}{lccc|ccc|ccc}
 & \multicolumn{3}{c|}{TREC-CAR} & \multicolumn{3}{c|}{Jeopardy} & \multicolumn{3}{c}{MSA}\\
Method & R@40 & P@10 & MAP & R@40 & P@10 & MAP & R@40 & P@10 & MAP \\
\noalign{\vskip 1mm}
\hline
\noalign{\vskip 1mm}
Raw-BM25 & 43.6 & 7.24 & 19.6 & 23.4 & 1.47 & 7.40 & 12.9 & 7.24 & 3.36 \\
Raw-Google & - & - & - & 30.1 & 1.92 & 7.71 & - & - & -\\
\noalign{\vskip 1mm}
\hline
\noalign{\vskip 1mm}
PRF-TFIDF & 44.3 & 7.31 & 19.9 & 29.9 & 1.91 & 7.65 & 13.2 & 7.27 & 3.50\\
PRF-RM & 45.1 & 7.35 & 19.5 & 30.5 & 1.96 & 7.64 & 12.3 & 7.22 & 3.38\\
PRF-Emb & 44.5 & 7.32 & 19.0 & 30.1 & 1.92 & 7.74& 12.2 & 7.22 & 3.20\\
Vocab-Emb & 44.2 & 7.30 & 19.1 & 29.4 & 1.87 & 7.80 & 12.0 & 7.21 & 3.21\\
\noalign{\vskip 1mm}
\hline
\noalign{\vskip 1mm}
SL-FF & 44.1 & 7.29 & 19.7 & 30.8 & 1.95 & 7.70 & 13.2 & 7.28 & 3.88\\
SL-CNN & 45.3 & 7.35 & 19.8 & 31.1 & 1.98 & 7.79 & 14.0 & 7.42 & 3.99\\
SL-Oracle & 50.8 & 8.25 & 21.0 & 38.8 & 2.50 & 9.92 & 17.3 & 10.12 & 4.89\\
\noalign{\vskip 1mm}
\hline
\noalign{\vskip 1mm}
RL-FF & 44.1 & 7.29 & 20.0 & 31.0 & 1.98 & 7.81 & 13.9 & 7.33 & 3.81\\
RL-CNN & 47.3 & 7.45 & 20.3 & 33.4 & \textbf{2.14} & 8.02 & 14.9 & 7.63 & 4.30\\
RL-RNN & \textbf{47.9}$^{*}$ & \textbf{7.52} & \textbf{20.6}$^{*}$  & \textbf{33.7}$^{*}$ & 2.12 & \textbf{8.07} & \textbf{15.1}$^{*}$ & \textbf{7.68} & \textbf{4.35}\\
RL-RNN-SEQ & 47.4 & 7.48 & 20.3  & 33.4 & 2.13 & 8.01 & 14.8 & 7.63 & 4.27\\
RL-Oracle & 55.9 & 9.06 & 23.0 & 42.4 & 2.74 & 10.3 & 24.6 & 12.83 & 6.33
\end{tabular}
\end{small}
\end{center}
\caption{Results on Test sets. We use R@40 as a reward to the RL-based models. We show the best results in bold. We use $*$ to denote statistically significant results ($p < 0.05$) against SL-CNN and PRF-RM baselines according to Student's paired t-test with a Bonferroni correction (code modified from \url{https://github.com/castorini/Anserini/blob/master/src/main/python/compare_runs.py})}
\label{tab:query_reformulation_results}

\end{table*}

\subsection{Metrics and Reward}

Three metrics are used to evaluate effectiveness:

\subparagraph{Recall@K:} Recall of the top-K retrieved documents: 
\begin{equation}
	\text{R}@K = \frac{|D_K \cap D^*|}{|D^*|},
\end{equation}
where $D_K$ are the top-$K$ retrieved documents and $D^*$ are the relevant documents. Since one of the goals of query reformulation is to increase the proportion of relevant documents returned, recall is our main metric.

\subparagraph{Precision@K:} Precision of the top-K retrieved documents:
\begin{equation}
	\text{P}@K = \frac{|D_K \cap D^*|}{|D_K|}
\end{equation}
Precision captures the proportion of relevant documents among the returned ones. Despite not being the main goal of a reformulation method, improvements in precision are also expected with a good query reformulation method. Therefore, we include this metric.

\subparagraph{Mean Average Precision:} The average precision is defined as:
\begin{equation}
	\text{AP} = \frac{\sum_{k} \text{P}@k \times \text{rel}(k)}{|D^*|},
\end{equation}
where
\begin{equation}
	\text{rel}(k) = 
    \begin{cases}
        1, & \text{if the k-th document is relevant;}\\
        0, & \text{otherwise.}
    \end{cases}
\end{equation}
The mean average precision of a set of queries $Q$ is then:
\begin{equation}
	\text{MAP} = \frac{1}{|Q|}\sum_{q \in Q} \text{AP}_q,
\end{equation}
where $\text{AP}_q$ is the average precision for a query $q$. This metric values the position of a relevant document in a returned list and is, therefore, complementary to precision and recall. 


\subparagraph{Reward:}
We use $\text{R}@K$ as a reward when training the proposed RL-based models as this metric has shown to be effective in improving the other metrics as well.

\subparagraph{SL-Oracle:}
In addition to the baseline methods and the proposed reinforcement learning approach, we report two oracle effectiveness bounds. The first oracle is a supervised learning oracle (SL-Oracle). It is a classifier that perfectly selects terms that will increase effectiveness according to the procedure described in Section~\ref{sec:proposed_methods}. This measure serves as an upper-bound for the supervised methods. Notice that this heuristic assumes that each term contributes independently from all the other terms to the retrieval effectiveness. There may be, however, other ways to explore the dependency of terms that would lead to higher effectiveness.

\subparagraph{RL-Oracle:}
Second, we introduce a reinforcement learning oracle (RL-Oracle) which estimates a conservative upper-bound effectiveness for the RL models. Unlike the SL-Oracle, it does not assume that each term contributes independently to the retrieval effectiveness. It works as follows: first, the \textit{validation} or \textit{test} set is divided into $N$ small subsets $\{A_i\}_{i=1}^N$ (each with 100 examples, for instance).
An RL model is trained on each subset $A_i$ until it overfits, that is, until the reward $R_i^*$ stops increasing or an early stop mechanism ends training.\footnote{The subset should be small enough, or the model should be large enough so it can overfit.} Finally, we compute the oracle effectiveness $R^*$ as the average reward over all the subsets: $R^*= \frac{1}{N}\sum_{i=1}^{N} R_i^*$.

This upper bound by the RL-Oracle is, however, conservative since there might exist better reformulation strategies that the RL model was not able to discover.



\begin{table}
\begin{center}
\begin{tabular}{lccc}
& TREC-CAR & Jeopardy & MSA \\
\noalign{\vskip 1mm}
\hline
\noalign{\vskip 1mm}
SL-Oracle & 13\% & 5\% & 11\% \\
RL-Oracle & 29\% & 27\% & 31\% \\
\end{tabular}
\end{center}
\caption{Percentage of relevant terms over all the candidate terms according to SL- and RL-Oracle.}

\label{tab:oracleterms}
\end{table}

\subsection{Implementation Details}

\subparagraph{Search engine:} We use Lucene as the search engine and BM25 as the ranking function for all PRF, SL, and RL methods. For Raw-Google, we restrict the search to the \textit{wikipedia.org} domain when evaluating its effectiveness on the Jeopardy dataset. We could not apply the same restriction to the two other datasets as Google does not index Wikipedia paragraphs, and as it is not trivial to match papers from MS Academic to the ones returned by Google Search.

\begin{figure}
\begin{center}
\centerline{\includegraphics[width=0.8\columnwidth]{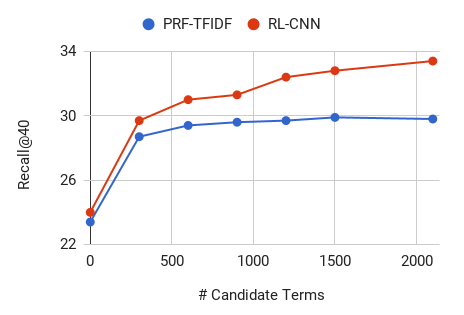}}
\caption{Our RL-based model continues to improve recall as more candidate terms are added, whereas a classical PRF method saturates.
}
\label{fig:feedback_terms}
\end{center}

\end{figure}

\subparagraph{Candidate terms:}
Inspired by~\citet{diaz2006improving}, we use Wikipedia articles as a source for candidate terms since it is a well-curated, clean corpus, with diverse topics.

At training and test times of SL methods, and at test time of RL methods, the candidate terms are from the first $M$ words of the top-$K$ Wikipedia articles retrieved. We select $M$ and $K$ using grid search on the validation set over $\{50,100,200,300\}$ and $\{1,3,5,7\}$, respectively. The best values are $M=300$ and $K=7$. These correspond to the maximum number of terms we could fit in a single 12 GB GPU.

At training time of an RL model, we use only \textit{one} document uniformly sampled from the top-$K$ retrieved ones as a source for candidate terms, as this leads to faster learning.

For the PRF methods, the top-$M$ terms according to a relevance metric (i.e., TF-IDF for PRF-TFIDF, cosine similarity for PRF-Emb, and conditional probability for PRF-RM) from each of the top-$K$ retrieved documents are added to the original query. We select $M$ and $K$ using grid search over $\{10, 50, 100, 200, 300, 500\}$ and $\{1, 3, 5, 9, 11\}$, respectively. The best values are $M=300$ and $K=9$.

\subparagraph{Multiple Reformulation Rounds:} Although our framework supports multiple rounds of search and reformulation, we did not find any significant improvement in reformulating a query more than once.
Therefore, the numbers reported in the results section were all obtained from models running two rounds of search and reformulation.

\subparagraph{Neural Network Setup:} 
For SL-CNN and RL-CNN variants, we use a 2-layer convolutional network for the original query. Each layer has a window size of 3 and 256 filters. We use a 2-layer convolutional network for candidate terms with window sizes of 9 and 3, respectively, and 256 filters in each layer. We set the dimension $d$ of the weight matrices $W,S,U$, and $V$ to $256$. For the optimizer, we use ADAM~\cite{kingma2014adam} with $\alpha=10^{-4}$, $\beta_1=0.9$, $\beta_2=0.999$, and $\epsilon=10^{-8}$. We set the entropy regularization coefficient $\lambda$ to $10^{-3}$.

For RL-RNN and RL-RNN-SEQ, we use a 2-layer bidirectional LSTM with 256 hidden units in each layer. We clip the gradients to unit norm. For RL-RNN-SEQ, we set the maximum possible number of generated terms to 50, and we use beam search of size four at test time.

We fix the dictionary of pretrained word embeddings during training, except the vector for out-of-vocabulary words. We found that this led to faster convergence and observed no difference in the overall effectiveness when compared to learning embeddings during training.

\section{Results and Discussion}

Table~\ref{tab:query_reformulation_results} shows the main result. As expected, reformulation based methods work better than using the original query alone. Supervised methods (SL-FF and SL-CNN) have in general better effectiveness than unsupervised ones (PRF-TFIDF, PRF-RM, PRF-Emb, and Emb-Vocab), but perform worse than RL-based models (RL-FF, RL-CNN, RL-RNN, and RL-RNN-SEQ).

RL-RNN-SEQ performs slightly worse than RL-RNN but produces queries that are three times shorter, on average (15 vs. 47 words). Thus, RL-RNN-SEQ is faster in retrieving documents and therefore might be a better candidate for a production implementation.

The effectiveness gap between the oracle and best performing method (Table~\ref{tab:query_reformulation_results}, RL-Oracle vs. RL-RNN) suggests that there is a large room for improvement. The cause for this gap is unknown, but we suspect, for instance, an inherent difficulty in learning a good selection strategy and the partial observability from using a black-box search engine.

\subsection{Relevant Terms per Document}

The proportion of relevant terms selected by the SL- and RL-Oracles over the total number of candidate terms (Table~\ref{tab:oracleterms}) indicates that only a small subset of terms are useful for the reformulation. Thus, we may conclude that the proposed method was able to learn an effective term selection strategy in an environment where relevant terms are infrequent.

\subsection{Scalability: Number of Terms vs Recall}

Figure~\ref{fig:feedback_terms} shows the improvement in recall as more candidate terms are provided to a reformulation method. The RL-based model benefits from more candidate terms, whereas the classical PRF method quickly saturates. In our experiments, the best performing RL-based model uses the maximum number of candidate terms that we could fit on a single GPU. We, therefore, expect further improvements with more computational resources.

\begin{figure*}
\begin{center}
\centerline{\includegraphics[width=0.68\paperwidth]{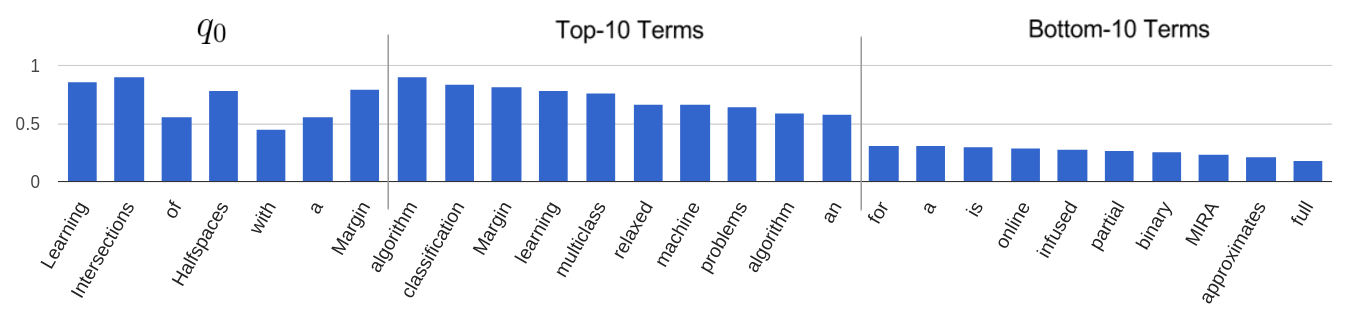}}
\vskip 0.1in
\centerline{\includegraphics[width=0.73\paperwidth]{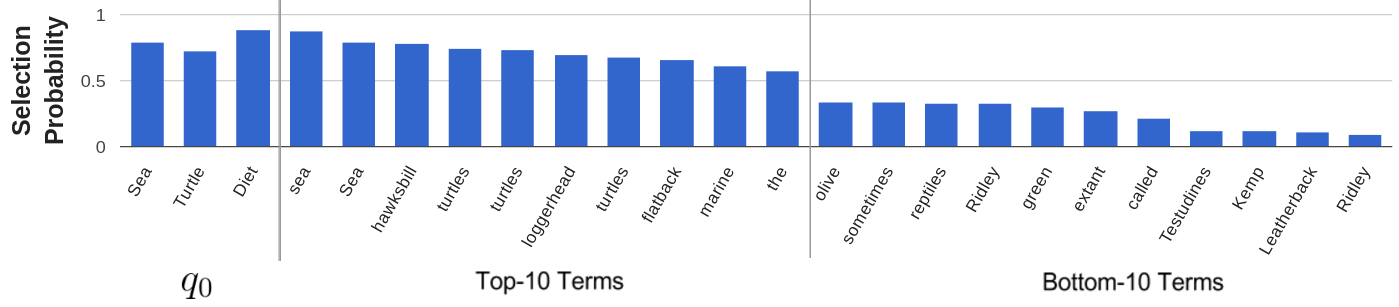}}
\caption{Probabilities assigned by the RL-CNN to candidate terms of two sample queries: ``\textit{Learning Intersections of Halfspaces with a Margin}'' (top) and ``\textit{Sea Turtle Diet}'' (bottom). We show the original query terms and the top-10 and bottom-10 document terms with respect to their probabilities. 
}
\label{fig:probs}
\end{center}
\end{figure*}

\begin{table}
\begin{center}
\begin{footnotesize}
\begin{tabular}{l|l}
Query & Top-3 Retrieved Documents \\
\noalign{\vskip 1mm}
\hline
\noalign{\vskip 1mm}
(Original) \textit{The Cross} & \textit{-The Cross Entropy Method for Network Reliability Estim.} \\
\textit{Entropy Method for} & \textit{\textbf{-Robot Weightlifting by Direct Policy Search}}\\
\textit{Fast Policy Search} & \textit{-Off-policy Policy Search}\\
\noalign{\vskip 1mm}
\hline
\noalign{\vskip 1mm}
(Reformulated) \textit{Cross } & \textbf{\textit{-Near Optimal Reinforcement}}\\
\textit{Entropy Fast Policy } & \textbf{\textit{Learning in Polynom. Time}}\\
\textit{Reinforcement } & \textit{-The Cross Entropy Method}\\
\textit{Learning policies} & \textit{for Network Reliability Estim.}\\
\textit{global search}& \textit{\textbf{-Robot Weightlifting by Direct Policy Search}}\\
\textit{optimization biased} & \textit{}\\
\noalign{\vskip 2mm}
\hline
\noalign{\vskip 1mm}
\hline
\noalign{\vskip 1mm}
(Original) \textit{Daikon} & ``\textit{...many types of pickles are made with daikon, includ...}''\\
\textit{Cultivation} & ``\textbf{\textit{...varieties of daikon can be grown as...}}''\\
& ``\textit{In Chinese cuisine, turnip cake and chai tow kway...}''\\
\noalign{\vskip 1mm}
\hline
\noalign{\vskip 1mm}
(Reformulated) \textit{Daikon} & ``\textit{...many types of pickles are made with daikon, includ...}''\\
\textit{Cultivation root seed } & \\
\textit{grow fast-growing} & ``\textbf{\textit{Certain varieties of daikon can be grown as a winter...}}''\\
\textit{Chinese leaves}& ``\textbf{\textit{The Chinese and Indian varieties tolerate higher...}}\\
\end{tabular}
\end{footnotesize}
\end{center}
\caption{Top-3 retrieved documents using the original query and a query reformulated by our RL-CNN model. In the first example, we only show the titles of the retrieved MSA papers. In the second example, we only show some words of the retrieved TREC-CAR paragraphs. \textbf{Bold} corresponds to relevant documents.}
\label{tab:return_example}

\end{table}

\begin{table}
\begin{center}
\begin{small}
\begin{tabular}{l|l}
Trained on & Selected Terms\\
\noalign{\vskip 1mm}
\hline
\noalign{\vskip 1mm}
TREC-CAR & \textit{serves american national Winsted accreditation}\\
\noalign{\vskip 1mm}
\hline
\noalign{\vskip 1mm}
Jeopardy & \textit{Tunxis Quinebaug Winsted NCCC}\\
\noalign{\vskip 1mm}
\hline
\noalign{\vskip 1mm}
MSA & \textit{hospital library arts center cancer center summer programs}\\
\end{tabular}
\end{small}
\end{center}
\caption{Given the query \textit{``Northwestern Connecticut Community College''}, models trained on different tasks choose different terms.}
\label{tab:samequery}
\end{table}




\subsection {Qualitative Analysis}

We show two examples of queries and the probabilities of each candidate term of being selected by the RL-CNN model in Figure~\ref{fig:probs}.

Notice that terms that are more related to the query have higher probabilities, although common words such as ``\textit{the}'' are also selected. This is a consequence of our choice of a reward that does not penalize the selection of neutral terms.


In Table~\ref{tab:return_example}, we show an original and reformulated query examples extracted from the MS Academic and TREC-CAR datasets, and their top-3 retrieved documents. Notice that the reformulated query retrieves more relevant documents than the original one. As we conjectured earlier, we see that a search engine tends to return a document simply with the largest overlap in the text, necessitating the reformulation of a query to retrieve semantically relevant documents.

\subparagraph{Same query, different tasks:} We compare in Table~\ref{tab:samequery} the reformulation of a sample query made by models trained on different datasets. The model trained on TREC-CAR selects terms that are similar to the ones in the original query, such as ``\textit{serves}'' and ``\textit{accreditation}''. These selections are expected for this task since similar terms can be effective in retrieving similar paragraphs. On the other hand, the model trained on Jeopardy prefers to select proper nouns, such as ``\textit{Tunxis}'', as these have a higher chance of being an answer to the question. The model trained on MSA selects terms that cover different aspects of the entity being queried, such as ``\textit{arts center}'' and ``\textit{library}'', since retrieving a diverse set of documents is necessary for the task the of citation recommendation.

\subsection {Training and Inference Times}
Our best model, RL-RNN, takes 8-10 days to train on a single K80 GPU. At inference time, it takes approximately one second to reformulate a batch of 64 queries. Approximately 40\% of this time is to retrieve documents from the search engine.

\section{Scaling Up Query Reformulation}
\label{sec:scaling_up_query_reformulation}

In the previous sections, we presented an agent that learns how to use a search engine by rewriting the original queries. The proposed agent, however, can be improved in two aspects. First, the oracle effectiveness showed that there exist reformulations that would lead to better effectiveness, but the agents were not able to learn how to produce them. \citet{aqa-iclr:2018} report a similar observation. The other potential improvement is training time. For example, the best agent trains in 10 days on a single GPU. There are effective methods to parallelize training using multiple GPUs, but these require careful implementation of how each machine exchange their weights or gradients. 

To improve our reformulation method, we are motivated by the observation that in reinforcement learning efficient exploration is key to achieve good effectiveness. The ability to explore in parallel a diverse set of strategies often speeds up training and leads to a better policy~\cite{mnih2016asynchronous,osband2016deep}. 

In the second half of this chapter, we propose a simple method to achieve efficient parallelized exploration of diverse policies, inspired by hierarchical reinforcement learning~\cite{singh1992reinforcement,lin1993reinforcement,
  dietterich2000hierarchical, dayan1993feudal}. 
We structure the agent into multiple \textit{sub-agents}, which are trained on disjoint subsets of the training data.
Sub-agents are co-ordinated by a meta-agent, called \textit{aggregator}, that groups and scores answers from the sub-agents for each given input. Unlike sub-agents, the aggregator is a generalist since it learns a policy for the entire training set. 

We argue that it is easier to train multiple sub-agents than a single generalist one since each sub-agent only needs to learn a policy that performs well for a subset of examples. Moreover, specializing agents on different partitions of the data encourages them to learn distinct policies, thus giving the aggregator the possibility to see answers from a population of diverse agents. Learning a single policy that results in an equally diverse strategy is more challenging.  

Since each sub-agent is trained on a fraction of the data, and there is no communication between them, training can be done faster than training a single agent on the full data. Additionally, it is easier to parallelize than applying existing distributed algorithms such as asynchronous SGD or A3C~\cite{mnih2016asynchronous}, as the sub-agents do not need to exchange weights or gradients. After training the sub-agents, only their actions need to be sent to the aggregator.

We show that it outperforms a strong baseline of an ensemble of agents trained on the full dataset. We also found that effectiveness and reformulation diversity are correlated.  

Our main contributions for the remaining of this chapter are the following:
\begin{itemize}
\setlength\itemsep{1pt}
\item A simple method to achieve more diverse strategies and better generalization effectiveness than a model average ensemble.
\item Training can be easily parallelized in the proposed method.
\item An interesting finding that contradicts our, perhaps naive,
  intuition: specializing agents on semantically similar data does not
  work as well as random partitioning. 
\end{itemize}

\subsection{Related Work}

The proposed approach is inspired by the mixture of experts, which was
introduced more than two decades
ago~\cite{jacobs1991adaptive,jordan1994hierarchical} and has been a
topic of intense study since then. The idea consists of training a
set of agents, each specializing in some task or data. One or
more gating mechanisms then select subsets of the agents that will
handle a new input. Recently, \citet{shazeer2017outrageously}
revisited the idea and showed strong effectiveness in the supervised
learning tasks of language modeling and machine translation. Their
method requires that output vectors of experts are exchanged between
machines. Since these vectors can be large, the network bandwidth
becomes a bottleneck. They used a variety of techniques to mitigate
this problem.~\citet{anil2018large} later proposed a method to further
reduce communication overhead by only exchanging the probability distributions of the different agents. Our method, instead, requires only scalars (rewards) and short strings (original query, reformulations, and answers) to be exchanged. Therefore, the communication overhead is small. 

Previous works used specialized agents to improve exploration in
RL~\cite{dayan1993feudal,singh1992reinforcement,kaelbling1996reinforcement}. For
instance, \citet{stanton2016curiosity} and \citet{conti2017improving}
use a population of agents to achieve a high diversity of strategies
that leads to better generalization effectiveness and faster
convergence. ~\citet{rusu2015policy} use experts to learn subtasks and
later merge them into a single agent using
distillation~\cite{hinton2015distilling}. 

The experiments are often carried out in simulated environments, such
as robot control~\cite{brockman2016openai} and
video-games~\cite{bellemare2013arcade}. In these environments, rewards
are frequently available, the states have low diversity (e.g., same
image background), and responses usually are fast (60 frames per
second). We, instead, evaluate our approach on tasks whose inputs
(queries) and states (documents and answers) are diverse because they
are in natural language, and the environment responses are slow (0.5-5
seconds per query). 

Somewhat similarly motivated is the work
of~\citet{serban2017deep}. They train many heterogeneous response
models and further train an RL agent to pick one response per
utterance. 

\begin{figure*}[ht]
\begin{center}
\centerline{\includegraphics[width=\textwidth]{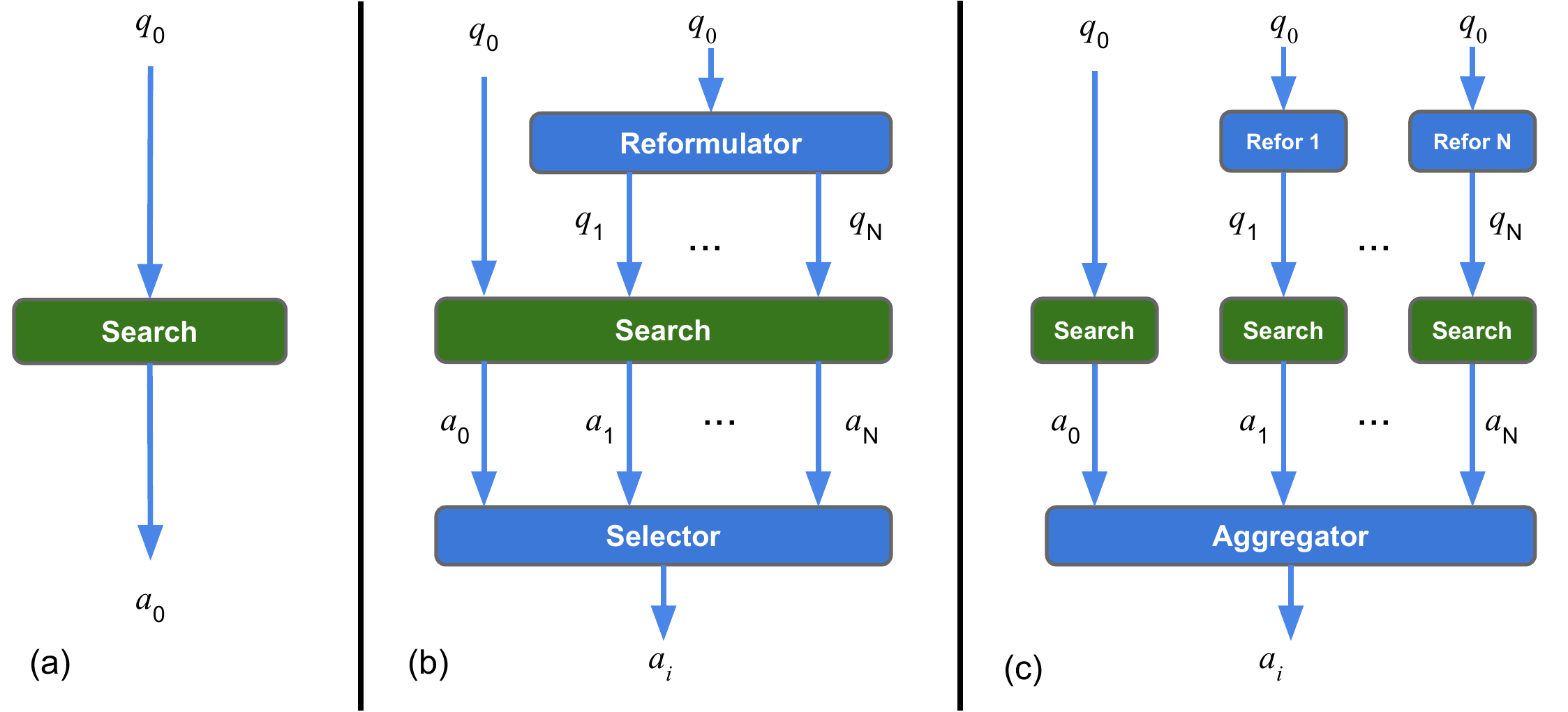}}
\caption{\textbf{a)} A vanilla search system. The
  query $q_0$ is given to the system which outputs a result
  $a_0$. \textbf{b)} The search system with a reformulator. The
  reformulator queries the system with $q_0$ and its
  reformulations $\{q_1, ... q_N\}$, and receives back the results
  $\{a_0, ..., a_N\}$. A selector then decides the best result $a_i$ for
  $q_0$. \textbf{c)} The proposed system. The original query is
  reformulated multiple 
  times by different reformulators. Reformulations are used to
  obtain results from the search system, which are then sent to the
  aggregator, which picks the best result for the original query
  based on a learned weighted majority voting scheme. Reformulators
  are independently trained on disjoint partitions of the 
  dataset, thus increasing the variability of reformulations.} 
\label{fig:overview}
\end{center}
\end{figure*}

\subsection{Method}
Figure~\ref{fig:overview}-(c) illustrates the new agent. An input
query $q_0$ is given to the $N$ sub-agents. A sub-agent is any system that
accepts as input a query and returns a corresponding
reformulation. Thus, sub-agents can be 
heterogeneous. 

Here we train each sub-agent on a partition of the training set. The $i$-th
agent queries the underlying search system with reformulation $q_i$
and receives a result $a_i$. The set $\{(q_i, a_i) | 0 \leq i \leq N\}$ is
given to the aggregator, which then decides which result will be final.


\paragraph{Sub-agents:}

The first step for training the new agent is to partition the training
set. We randomly split it into equal-sized subsets. 
In our implementation, a sub-agent is a sequence-to-sequence
model~\cite{sutskever2014sequence,cho2014learning} trained on a
partition of the dataset. It receives as an input the original
query $q_0$ and outputs a list of reformulated queries $(q_i)$
using beam search. 

Each reformulation $q_i$ is given to the same environment that returns
a list of results $(a_i^1, .., a_i^K)$ and their respective rewards
$(r_i^1,.. r_i^K)$. We then use REINFORCE~\cite{williams1992simple} to
train the sub-agent. At training time, instead of using beam search,
we sample reformulations.  

Note that we also add the identity agent (i.e., the reformulation is
the original query) to the pool of sub-agents.  

\paragraph{Aggregator:}

The aggregator receives as inputs $q_0$ and a list of candidate results
$(a_i^1,.. a_i^K)$ for each reformulation $q_i$. We first compute the
set of unique results ${a_j}$ and two different scores for each
result: the accumulated rank score $s_j^A$ and the relevance score $s_j^R$.  

The accumulated rank score is computed as $s_j^A = \sum_{i=1}^N
\frac{1}{\text{rank}_{i,j}}$, where $\text{rank}_{i,j}$ is the rank of the j-th
result when retrieved using $q_i$. The relevance score $s_j^R$ is the prediction that the result $a_j$ is
relevant to query $q_0$. It is computed as: 
\begin{equation}
s_j^R = \sigma(W_2 \text{ReLU}(W_1 z_j + b_1) + b_2),
\end{equation}
where
\begin{equation}
z_j = [f_{\text{CNN}}(q_0); f_{\text{BOW}}(a_j); f_{\text{CNN}}(q_0) - f_{\text{BOW}}(a_j); f_{\text{CNN}}(q_0) \odot f_{\text{BOW}}(a_j)], 
\label{eq:zi}
\end{equation}
$W_1 \in \RR^{4D \times D}$ and $W_2 \in \RR^{D \times 1}$ are weight
matrices, $b_1 \in \RR^{D}$ and $b_2 \in \RR^{1}$ are biases. The brackets in $[x; y]$ represent the concatenation of vectors $x$ and $y$. The symbol $\odot$ denotes the element-wise multiplication, $\sigma$ is the sigmoid function, and ReLU is a Rectified Linear
Unit function~\cite{nair2010rectified}. The function $f_{\text{CNN}}$ is
implemented as a CNN encoder\footnote{In the preliminary experiments, we found CNNs to work better than LSTMs~\cite{hochreiter1997long}.}
followed by average pooling over the
sequence~\cite{kim2014convolutional}. The function $f_{\text{BOW}}$ is the average word
embeddings of the result. At test time, the top-K answers with respect to $s_j={s_j^A s_j^R}$ are returned. 

We train the aggregator with stochastic gradient descent (SGD) to
minimize the cross-entropy loss: 
\begin{equation} \label{eq:aggregator_loss}
L = -\sum_{j \in J^*} \log (s_j^R) - \sum_{j \notin J^*} \log (1 - s_j^R),
\end{equation}
where $J^*$ is the set of indexes of the relevant results.

\paragraph{Hyperparameters:}

For the sub-agents, we use mini-batches of size 256, ADAM as the optimizer, and learning rate of $10^{-4}$.
For the aggregator, the encoder $f_{q_0}$ is a word-level two-layer CNN with filter sizes of 9 and 3, respectively, and 128 and 256 kernels, respectively. $D=512$. No dropout is used. ADAM is the optimizer with learning rate of $10^{-4}$ and mini-batch of size 64. It is trained for 100 epochs.

\subsection{Document Retrieval}

We now present experiments and results in a document retrieval task. In this task, the goal is to rewrite a query so that the number of relevant documents retrieved by a search engine increases.

The environment and datasets are the same to ones we used to evaluate the single-agent query reformulator, except for the TREC-CAR dataset, in which we use different training, validation and test set splits, namely, we use the first four folds for training, the last fold for validation and the automatic annotations benchmark of 2017 (approx. 1,800 queries) for test.


\begin{table*}
\begin{center}
\begin{footnotesize}
\begin{tabular}{l|ccc|cc}
 & TREC-CAR & Jeopardy & MSA & \multicolumn{2}{c}{Training Cost}\\
 & & & & Days & FLOPs ($\times 10^{18}$)\\
\noalign{\vskip 1mm}
\hline
\noalign{\vskip 1mm}
BM25 & 12.3 & 8.2 & 3.1 & - & - \\
RM3 & 13.0 & 13.5 & 3.1 & - & - \\

\noalign{\vskip 1mm}
\hline
\noalign{\vskip 1mm}
RL-RNN & 13.8 & 15.9 & 4.1 &10 & 2.3\\
RL-10-Ensemble & 14.1 & 16.2 & 4.3 & 10 & 23.0\\
\noalign{\vskip 1mm}
\hline
\noalign{\vskip 1mm}
RL-10-Full & 15.1 & 17.0 & 4.4 & 1 & 2.3\\
RL-10-Bagging & 15.3 & 17.2 & 4.6 & 1 & 2.3\\
RL-10-Sub & 15.7 & 17.4 & 4.5 & 1 & 2.3\\
RL-10-Sub (Pretrained$^{\star}$) & 16.0 & 17.5 & 4.6 & 10$^{\star}$+1 & 4.6\\
RL-10-Full (Extra Budget) & \textbf{16.2} & \textbf{17.9} & \textbf{4.7} & 10 & 23.0\\
\end{tabular}
\end{footnotesize}
\end{center}
\vskip -1mm
\caption{MAP on the test set of the document retrieval datasets. $^{\star}$The weights of the agents are initialized from a single model pretrained on the full training set for 10 days.}
\label{tab:results_document_retrieval}
\end{table*}

\paragraph{Baselines:} We use the following methods as baselines:

\subparagraph{BM25:} We give the original query to Lucene in its default configuration with BM25 as the ranking function, and use the retrieved documents as results.

\subparagraph{RM3:}  We reimplement the relevance model for query expansion of~\citet{lavrenko2001relevance} with a Dirichlet smoothed language model~\cite{zhai2001study}, and use the top-$N$ terms with highest posterior $P(t|q_0)$ as the new query.


\subparagraph{RL-RNN:} This is the sequence-to-sequence model trained with reinforcement learning from Section~\ref{sec:proposed_methods}. The reformulated query is formed by appending new terms to the original query. The terms are selected from the documents retrieved using the original query. This agent is trained from scratch. 

\subparagraph{RL-N-Ensemble:} We train $N$ RL-RNN agents with different initial weights on the full training set. At test time, we average the probability distributions of all the $N$ agents at each time step and select the token with the highest probability, as done by~\citet{sutskever2014sequence}.

\paragraph{Proposed Methods:} We evaluate the following methods:

\subparagraph{RL-N-Full:} We train $N$ RL-RNN agents with different
initial weights on the full training set. The answers are obtained
using the best (greedy) reformulations of all the agents and are given
to the aggregator.

\subparagraph{RL-N-Bagging:} This is the same as RL-N-Full but we construct the training set of each RL-RNN agent by sampling with replacement D times from the full training set, which has a size of D. This is known as the bootstrap sample and leads to approximately 63\% unique samples, the rest being duplicates. Note that this not exactly the bagging method~\cite{breiman1996bagging} because our aggregator is different from the average model prediction proposed in that work.

\subparagraph{RL-N-Sub:} This is the proposed agent. It is similar to RL-N-Full but the multiple sub-agents are trained on random partitions of the
dataset (see Figure~\ref{fig:overview}-(c)).

\paragraph{Results:}

A summary of the document retrieval results is shown in
Table~\ref{tab:results_document_retrieval}.
The proposed methods (RL-10-\{Sub, Bagging, Full\}) have 20-60\% relative
retrieval improvement over the standard ensemble (RL-10-Ensemble)
while training ten times faster. More interestingly, RL-10-Sub has a better retrieval than the single-agent version (RL-RNN), uses the same computational budget, and trains on a fraction of the time. Lastly, we found that RL-10-Sub (Pretrained) has the best balance between effectiveness and training cost across all datasets.

We estimate the number of floating point operations used to train a model by multiplying the training time, the number of GPUs used, and 2.7 TFLOPS as an estimate of the single-precision floating-point of a K80 GPU.

Since the sub-agents are frozen during the training of the aggregator, we pre-compute all $(q_0, q_i, a_i, r_i)$ tuples from the training set, thus avoiding sub-agent or environment calls. This reduces its training time to less than 6 hours ($0.06\times10^{18}$ FLOPs). Since this cost is negligible when compared to the sub-agents', we do not include it in the table.

\begin{figure*}
\begin{center}
\centerline{
  \includegraphics[width=0.32\textwidth]{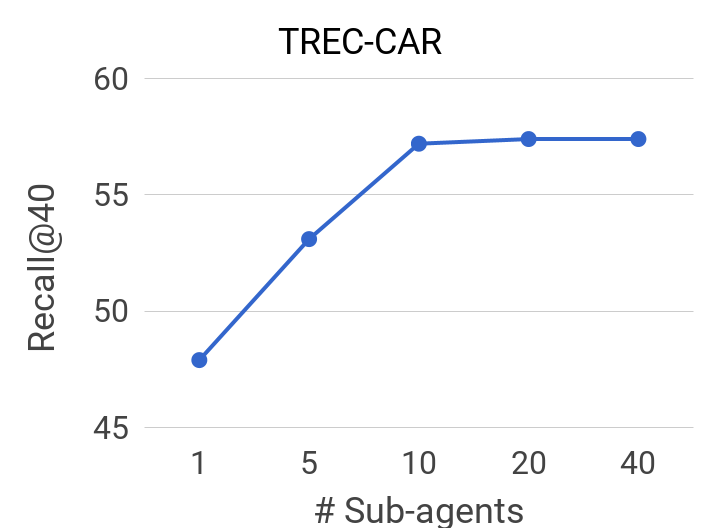}
  \includegraphics[width=0.3\textwidth]{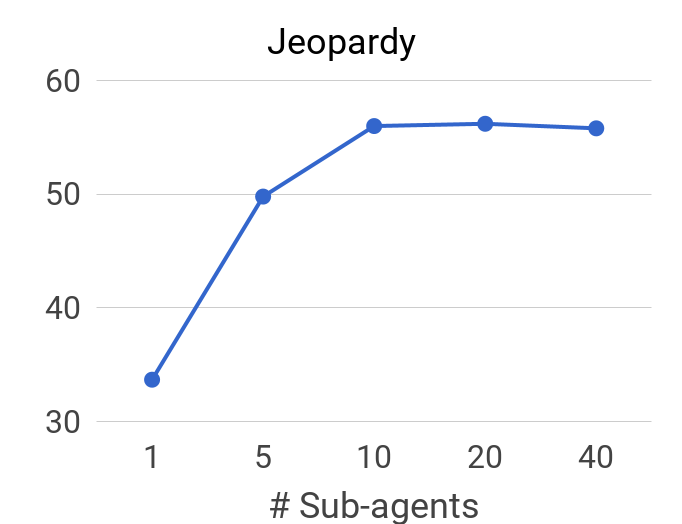}
  \includegraphics[width=0.3\textwidth]{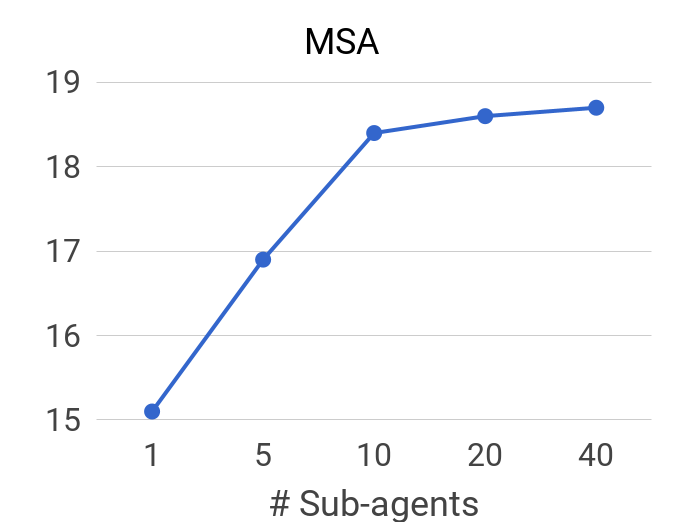}
}
\caption{Overall system's effectiveness for different number of sub-agents.}
\label{fig:number_sub_agents}
\end{center}
\end{figure*}

\subparagraph{Number of Sub-Agents:} We compare the retrieval effectiveness of the full system (reformulators + aggregator) for different numbers of agents in Figure~\ref{fig:number_sub_agents}. The effectiveness of the system is stable across all datasets after more than ten sub-agents are used, thus indicating the robustness of the proposed method.

\subparagraph{Comparison of Aggregator Functions:}
To validate the effectiveness of the proposed aggregation function, we conducted a comparison study on the TREC-CAR dataset. We present the results in Table~\ref{tab:comparison}. We notice that removing or changing the accumulated rank or relevance score functions results in a retrieval effectiveness drop between 2-8\%. 

We also experimented concatenating to the input vector $z_i$ (eq.~\ref{eq:zi}) a vector to represent each sub-agent. These vectors were learned during training and allow the aggregator to distinguish sub-agents. However, this did not lead to an improvement in effectiveness.

\subsection{Question Answering}

To further assess the effectiveness of the proposed method, we conduct
experiments in a question-answering task, comparing our agent with
the active question-answering agent proposed by~\citet{aqa-iclr:2018}.

\begin{table}
\begin{center}
\begin{tabular}{l|c}
Aggregator Function & Recall@40 \\
\noalign{\vskip 1mm}
\hline
$s_j = s_j^A s_j^R$ (reference) & 57.3 \\
$s_j = s_j^R$ & -5.0 \\
$s_j = s_j^A$ & -8.1 \\
$s_j^A = \sum_{i=1}^N \mathbbm{1}_{a_i=a_j}$ & -3.2 \\
$z_j=f_{\text{CNN}}(q_0)||f_{\text{BOW}}(a_j)$ (eq.~\ref{eq:zi}) & -2.4 \\
\end{tabular}
\end{center}
\caption{Comparison of different aggregator functions on TREC-CAR. The reference model (first row) is an RL-10-Sub.}
\label{tab:comparison}
\end{table}


The environment receives a question as an action and returns an
answer as an observation, and a reward computed against a
relevant answer. We use BiDAF as the question-answering 
system~\cite{seo2016bidirectional}. Given a
question, it outputs an answer span from a 
list of snippets. We use as a reward the token level F1 score on the
answer (we will define it precisely later on).  

We follow~\citet{aqa-iclr:2018} to train BiDAF. We
emphasize that BiDAF's parameters are frozen when we train and evaluate
the reformulation system. Training and evaluation are performed on the SearchQA
dataset~\cite{dunn2017searchqa}. The data contains \textit{Jeopardy!} clues
as questions. Each clue has a correct answer and a list of 50 snippets
from Google's top search results. The training, validation, and test
sets contain 99,820, 13,393, and 27,248 examples, respectively.

\paragraph{Baselines:} We compare our agent against the following baselines:

\subparagraph{BiDAF:} The original question is given to the
question-answering system without any modification (see
Figure~\ref{fig:overview}-(a)).

\subparagraph{Re-Ranker and R$^3$:} Re-Ranker is the best model
from~\citet{wang2017evidence}. They use an answer re-ranking approach
to reorder the answer candidates generated by a base Q\&A
model, R$^3$~\cite{wang2017r}. We report both systems' results as a
reference. To the best of our knowledge, they are currently the best systems on SearchQA. R$^3$
alone, without re-ranking, outperforms BiDAF by about 20 F1 points.

\subparagraph{AQA:} This is the best model from~\citet{aqa-iclr:2018}. It
consists of a reformulator and a selector. The reformulator is a
subword-based sequence-to-sequence model that produces twenty
reformulations of an input question using beam search. The
reformulations and their answers are given to the selector, which then
chooses one of the answers as final (see Figure~\ref{fig:overview}-(b)). The reformulator is pretrained on translation and paraphrase data. 

\begin{table*}
\begin{center}
\begin{small}
\begin{tabular}{l|cc|cc|cc}
 & \multicolumn{2}{c}{Dev} & \multicolumn{2}{|c|}{Test} & \multicolumn{2}{c}{Training}\\
 & F1 & Oracle & F1 & Oracle & Days & FLOPs ($\times 10^{18}$) \\
\noalign{\vskip 1mm}
\hline
\noalign{\vskip 1mm}
BiDAF~\cite{seo2016bidirectional} & 37.9 & - & 34.6 & - & - & - \\
R$^3$~\cite{wang2017r} & - & - & 55.3 & - & - & - \\
Re-Ranker~\cite{wang2017evidence} & - & - & 60.6 & - & - & -\\
\noalign{\vskip 1mm}
\hline
\noalign{\vskip 1mm}
AQA~\cite{aqa-iclr:2018} & 47.4 & 56.0 & 45.6 & 53.8 & 10 & 4.6 \\
\noalign{\vskip 1mm}
\hline
\noalign{\vskip 1mm}
AQA-10-Sub & 50.2 & 65.7 & 48.1 & 62.4 & 1 & 4.6 \\
AQA-10-Full & 49.6 & 60.1 & 47.5 & 57.8 & 1 & 4.6 \\
AQA-10-Full (extra budget) & 49.9 & 60.1 & 49.5 & 58.0 & 10 & 46.0 \\
\end{tabular}
\end{small}
\end{center}
\caption{Main result on the question-answering task (SearchQA dataset). We did not include the training cost of the aggregator (0.2 days, 0.06 $\times 10^{18}$ FLOPs).
}
\label{tab:results_question_answering}
\end{table*}

\paragraph{Proposed Methods:}

\subparagraph{AQA-N-\{Full, Sub\}:} Similar to the RL-N-\{Full, Sub\} models, we use AQA reformulators as the sub-agents followed by an aggregator to create AQA-N-Full and AQA-N-Sub models, whose sub-agents are trained on the full and random partitions of the dataset, respectively.

As for the hyperparameters of the sub-agents, we use mini-batches of size 64, SGD as the optimizer, and learning rate of $10^{-3}$. For the aggregator, the encoder $f_{q_0}$ is a token-level, three-layer CNN with filter sizes of 3, and 128, 256, and 256 kernels, respectively. We train it for 100 epochs with mini-batches of size 64 with SGD and learning rate of $10^{-3}$.

\paragraph{Evaluation Metrics:}

We use the macro-averaged F1 score as the main metric. It measures the average bag of tokens overlap between the prediction and relevant answer. We take the F1 over the relevant answer for a given question and then average over all of the questions. Additionally, we present the oracle effectiveness, which is from a perfect aggregator that predicts $s_j^R=1$ for relevant answers and $s_j^R=0$, otherwise.

\paragraph{Results:}

Results are presented in Table~\ref{tab:results_question_answering}. The proposed methods (AQA-10-\{Full,Sub\}) have both better F1 and oracle effectiveness than the single-agent AQA method while training in one-tenth of the time.\footnote{The 
original question and its answer are important contributors to the 
final effectiveness. If not given to the aggregator, the effectiveness of
all AQA models decreases by 1-2\% in F1.} Even when the ensemble method is given ten times more training time (AQA-10-Full, extra budget), our method achieves higher effectiveness.

The best model outperforms BiDAF, which is used in our environment, by almost 16 F1 points. In absolute terms, the proposed method does not reach the effectiveness of the Re-Ranker or underlying R$^3$ system. It is important to realize,
though, that these are orthogonal issues: any Q\&A system,
including R$^3$, could be used as environments, including re-ranking
post-processing. We leave this as a future work.

\paragraph{Query Diversity:} We also evaluate how query diversity and effectiveness are related, using the following metrics:

\subparagraph{pCos:} Mean pair-wise cosine distance:
\begin{equation}
\frac{1}{N} \sum_{n=1}^N \frac{1}{|Q^n|} \sum_{q, q' \in Q^n} \text{cosine}\big(\#q, \#q'\big),
\end{equation}
where $Q^n$ is a set of reformulated queries for the $n$-th original query in the development set and $\#q$ is the token count vector of q.

\subparagraph{pBLEU:} Mean pair-wise sentence-level BLEU~\cite{chen2014systematic}:
\begin{equation}
\frac{1}{N} \sum_{n=1}^N \frac{1}{|Q^n|} \sum_{q, q' \in Q^n} \text{BLEU}\big(q, q'\big)
\end{equation}

\subparagraph{PINC:} Mean pair-wise paraphrase in k-gram changes~\cite{chen2011collecting}:
\begin{equation}
\frac{1}{N} \sum_{n=1}^N \frac{1}{|Q^n|} \sum_{q, q' \in Q^n} \text{PINC}(q, q'),
\end{equation}
\begin{equation}
\text{PINC}(q, q') = \frac{1}{K} \sum_{k=1}^K 1 - \frac{|\text{k-gram}_q \cap \text{k-gram}_{q'}|}{|\text{k-gram}_{q'}|},
\end{equation}
where $K$ is the maximum number of k-grams considered (we use $K=4$).
\subparagraph{Length Std:} Standard deviation of the reformulation lengths: \begin{equation}
\frac{1}{N} \sum_{n=1}^N \text{std}\big(\{|q_i^n|\}_{i=1}^{|Q|}\big)
\end{equation}

Table~\ref{tab:diversity} shows that the multiple agents trained on partitions of the dataset (AQA-10-Sub) produce more diverse queries than a single agent with beam search (AQA) and multiple agents trained on the full training set (AQA-10-Full). This suggests that its higher effectiveness can be partly attributed to the higher diversity of the learned policies.

\begin{table*}
\begin{center}
\begin{tabular}{l|cccc|cc}
Method & pCos $\downarrow$ & pBLEU $\downarrow$ & PINC $\uparrow$ & Length Std $\uparrow$ & F1 $\uparrow$ & Oracle $\uparrow$\\
\noalign{\vskip 1mm}
\hline
\noalign{\vskip 1mm}
AQA & 66.4 & 45.7 & 58.7 & 3.8 & 50.7 & 60.0\\
AQA-10-Full & 29.5 & 26.6 & 79.5 & 9.2 & 52.9 & 63.1\\
AQA-10-Sub & \textbf{14.2} & \textbf{12.8} & \textbf{94.5} & \textbf{11.7} & \textbf{53.4} & \textbf{68.5} \\
\end{tabular}
\end{center}
\caption{Diversity scores of reformulations from different methods. For pBLEU and pCos, lower values mean higher diversity. Notice that higher diversity scores are associated with higher F1 and oracle scores.}
\label{tab:diversity}
\end{table*}

\paragraph{Training Stability of Single vs. Multi-Agent:} Reinforcement learning algorithms that use non- linear function approximators, such as neural networks, are known to be unstable \cite{tsitsiklis1996analysis, fairbank2011divergence, pirotta2013adaptive, mnih2015human}. Ensemble methods are known to reduce this variance~\cite{freund1995boosting,breiman1996bagging,breiman1996bias}. Since the proposed method can be viewed as an ensemble, we compare the AQA-10-Sub's F1 variance against a single agent (AQA) on ten runs. Our method has a much smaller variance: 0.20 vs. 1.07. We emphasize that it also has higher effectiveness than the AQA-10-Ensemble.
We argue that this higher stability is due to the use of multiple agents. Answers from agents that diverged during training can be discarded by the aggregator. In the single-agent case, answers come from only one, possibly bad, reformulation.

\paragraph{Reformulation Examples:} Table~\ref{tab:reformulation_examples} shows four reformulation examples from various methods. The proposed method (AQA-10-Sub) performs better in the first and second examples than the other methods. Note that, despite the large diversity of reformulations, BiDAF still returns the correct answer.
In the third example, the proposed method fails to produce the right answer, whereas the other methods perform well.
In the fourth example, despite the correct answer is in the set of returned answers, the aggregator fails to set a high score for it.

\begin{table*}[h!]
\begin{center}
\begin{tiny}
\begin{tabular}{r|p{7.2cm}|p{4.4cm}}
    \toprule
    {\bf Method} & {\bf Query} & {\bf Reference / Answer from BiDAF (F1)} \\
    \midrule
Jeopardy! & The name of this drink that can be blended or on the rocks means "daisy" in Spanish\\
SearchQA  & name drink blended rocks means daisy spanish & margarita \\ 
AQA & What name drink blended rocks mean daisy spanish? & margarita tequila daisy (0.33)\\
    & \textbf{What rock drink name means daisy spanish?} & \textbf{margarita tequila daisy mentioned (0.20)}\\
    & What name drink blended rocks means daisy spanish? & margarita tequila daisy mentioned (0.20)\\
    & What rock drinks name means daisy spanish? & margarita tequila daisy mentioned (0.20)\\
    & What name drink blended rock means daisy spanish? & margarita tequila daisy mentioned (0.20)\\
AQA-10-Full & What is drink name name drink daisy daisy? me & margarita eater jun (0.33) \\
    & What name is drink spanish? & margarita eater jun (0.33) \\
    & \textbf{What is daisy blender rock daisy spanish?? daisy spanish?} & \textbf{cocktail daisy margarita spanish (0.26)} \\
    & rock name name & cocktail daisy margarita spanish (0.25)\\
    & What name drink blended st st st st st ship ship & cocktail daisy margarita spanish (0.26)\\
AQA-10-Sub & Where is name drink?? & margarita (1.0)\\
    & \textbf{What is drink blended rock?} & \textbf{margarita (1.0)}\\
    & rock definition name & margarita (1.0) \\
    & What is name drink blended rock daisy spanish 16 daisy spanish? & margarita similarity (0.5) \\
    & Nam Nam Nam Nam Nam Nam Nam drink & tequila (0.0) \\
    \midrule
Jeopardy! & A graduate of Howard University, she won the Nobel Prize for literature in 1993 & \\
SearchQA & graduate howard university , nobel prize literature 1993 & toni morrison\\ 
AQA & Nobel university of howard university? & toni morrison american novelist (0.5)\\
    & Nobel university of howard university in 1993? & toni morrison american novelist (0.5)\\
    & \textbf{Nobel graduate literature in 1993?} & \textbf{toni morrison american novelist (0.5)}\\
    & Nobel university graduate howard university 1993? & princeton (0.0) \\
    & Nobel university for howard university? & columbia (0.0)\\
AQA-10-Full & Another university start howard university starther & toni morrison american novelist (0.5)\\
    & \textbf{university howard car?} & \textbf{toni morrison american novelist (0.5)}\\
    & What is howard graduate nobel? & toni morrison american novelist (0.5)\\
    & What is howard howard university? & toni morrison american novelist (0.5)\\
    & Where graduated howard university noble literature literature? & american novelist morrison (0.16) \\
AQA-10-Sub & \textbf{Where is howard university??} & \textbf{toni morrison (1.0)}\\
    & The nobel university? & toni morrison (1.0)\\
    & What name howard howard university? & toni morrison (1.0)\\
    & This howard? & toni morrison american novelist (0.5)\\
    & 1993? & howard cornell universities (0.0) \\
   \midrule
Jeopardy! & For Bill Gates, it computes to own 2 models, the 959 and the 911, from this manufacturer\\
SearchQA & bill gates , computes 2 models , 959 911 , manufacturer & porsche\\ 
AQA & \textbf{Bill gates iin computes older models? }& \textbf{porshe (1.0)} \\ 
    & Bill gates in compute gates how old are they? & porshe (1.0) \\
    & Bill gates bill gates computes mod? & porshe (1.0) \\
    & Bill gates computes 2 models pics of 959? & porshe (1.0) \\
    & Bill gates in compute gates how old is it? & porshe (1.0) \\
AQA-10-Full & Another model start bill bette & porshe (1.0) \\
    & What is an bill gates 100 car? & porshe (1.0) \\
    & What is bill bill bill bill gates computes? & porshe (1.0) \\
    & \textbf{What is manufacturer?} & \textbf{porshe (1.0)} \\
    & bill bill gats  sa computes 2 bill gats? & porshe (1.0)\\
AQA-10-Sub & Where is bill gates manufacturer? & bill gates (0.0)\\
    & \textbf{A bill gates?} & \textbf{bill gates (0.0)} \\
    & The model? & bill gates (0.0)\\ 
    & What is bill gates model? & sports car (0.0)\\
    & What model bill gates 9 58 model 9 gates? & sports car (0.0)\\
\midrule
Jeopardy! & The first written mention of this capital's name was in a 1459 document of Vlad the Impaler\\
SearchQA & first written mention capital 's name 1459 document vlad impaler & bucharest\\ 
AQA & First film was written by 1459 vlad impaler? & bucharest castle (0.5) \\
    & First film was written by 1459 vlad impalter? & bucharest castle (0.5)\\
    & First film was written by 1459 vlad impal? & bucharest castle (0.5) \\
    & First film was written by 1459 vlad impalot? & bucharest castle (0.5)\\
    & \textbf{First film was written in 1459?} & \textbf{bucharest national capital (0.33)} \\
AQA-10-Full & What is capital vlad impaler? & bucharest (1.0) \\
    & First referred capital vlad impaler impaler? & bucharest (1.0)\\
    & capital & romania 's largest city capital (0.0)\\
    & Another name start capital & romania 's largest city capital (0.0) \\
    & \textbf{capital capital vlad car capital car capital?} & \textbf{romania 's largest city capital (0.0)}\\
AQA-10-Sub & Where is vla capital capital vlad impalers? & bucharest (1.0) \\
    & What capital vlad capital document document impaler? & bucharest (1.0)\\
    & \textbf{Another capital give capital capital} & \textbf{bulgaria , hungary , romania (0.0)} \\
    & capital? & bulgaria , hungary , romania (0.0)\\
    & The name capital name? & hungary (0.0) \\
\bottomrule
\end{tabular}
\end{tiny}
\end{center}
\vspace{-4mm}
\caption{
Examples for the qualitative analysis on SearchQA.
In \textbf{bold} are the reformulations and answers that had the highest scores predicted by the aggregator. We only show the top-5 reformulations of each method. For a detailed analysis of the language learned by the reformulator agents, see~\citet{buck2018analyzing}.
}

\label{tab:reformulation_examples}
\end{table*}

\section{Summary}

In this chapter, we introduced a reinforcement learning framework for task-oriented automatic query reformulation. An appealing aspect of this framework is that an agent can be trained to use a search engine for a specific task. The empirical evaluation has confirmed that the proposed approach outperforms strong baselines in the three separate tasks. 

The analysis based on two oracle approaches has revealed that there is a meaningful room for further development, which motivated the development of the multiple agent framework. In this framework, we proposed a method to build a better query reformulation system by training multiple sub-agents on partitions of the data using reinforcement learning and an aggregator that learns to combine the answers of the multiple agents given a new query. We showed the effectiveness and efficiency of the proposed approach on the tasks of document retrieval and question answering.
\Chapter{Looking into the Black Box}{Introducing a Pretrained Re-ranker and a Novel Document Expansion Method}
\label{chap:blackbox}

Modern search engines are multi-stage pipelines, with query rewrite, initial index retrieval, and re-ranking being important components. 
In Chapter~\ref{chap:query_reformulation}, we focused on methods to improve a black-box search engine through a novel query rewriting agent.
Despite elegant, the separation of an agent and a black-box environment in the reinforcement learning framework might be a poor abstraction for many real-world applications.
For example, if we have access to the internal mechanism of the environment, we can alter it to achieve better performance.
In this chapter, we look inside the black box and modify two of its components: the re-ranker and inverted index. 

In the first half of this chapter, we describe in detail how we have re-purposed BERT as a passage re-ranker and achieved state-of-the-art results on two datasets.
In the second half, we will introduce a novel technique to augment documents prior to indexing and show that this enriched index results in better retrieval effectiveness.

\section{A Pretrained Re-ranker}
\label{sec:reranker}

We have seen rapid progress in machine reading compression in recent years with the introduction of large-scale datasets, such as SQuAD~\cite{rajpurkar2016squad}, MS MARCO~\cite{nguyen2016ms}, SearchQA~\cite{dunn2017searchqa}, TriviaQA~\cite{joshi2017triviaqa}, and QUASAR-T~\cite{dhingra2017quasar}, and the broad adoption of neural models, such as BiDAF~\cite{seo2016bidirectional}, DrQA~\cite{chen2017reading}, DocumentQA~\cite{clark2017simple}, and QAnet~\cite{yu2018qanet}.

The information retrieval (IR) community has also experienced a flourishing development of neural ranking models, such as DRMM~\cite{guo2016deep}, KNRM~\cite{xiong2017end}, Co-PACRR~\cite{hui2018co}, and DUET~\cite{mitra2017learning}. However, until recently, there were only a few large datasets for passage ranking, with the notable exception of the TREC-CAR~\cite{dietz2017trec}. This, at least in part, prevented the neural ranking models from being successful when compared to more classical IR techniques~\cite{lin2019neural}.

We argue that the same two ingredients that made possible much progress on the reading comprehension task are now available for the ranking task. Namely, the MS MARCO passage ranking dataset, which contains one million queries from real users and their respective relevant passages annotated by humans, and BERT~\cite{devlin2018bert}, a powerful general-purpose natural language processing model.

\subsection{Method}

The job of the re-ranker is to estimate a score $s_i$ of how relevant a candidate passage $d_i$ is to a query $q$.
We use BERT as our re-ranker. Using the same notation used by~\citet{devlin2018bert}, we feed the query as sentence A and the passage text as sentence B. We truncate the query to have at most 64 tokens. We also truncate the passage text such that the concatenation of query, passage, and separator tokens have a maximum length of 512 tokens. We use a $\text{BERT}_\text{LARGE}$ model as a binary classification model, that is, we use the $[\text{CLS}]$ vector as input to a single layer neural network to obtain the probability of the passage being relevant. We compute this probability for each passage independently and obtain the final list of passages by ranking them with respect to these probabilities. 

We start training from a pretrained BERT model and fine-tune it to our re-ranking task using the cross-entropy loss: 
\begin{equation} 
\label{eq:reranker_loss}
L = -\sum_{j \in J_{\text{pos}}} \log (s_j) - \sum_{j \in J_{\text{neg}}} \log (1 - s_j),
\end{equation}
where $J_{\text{pos}}$ is the set of indexes of the relevant passages and $J_{\text{neg}}$ is the set of indexes of non-relevant passages in top-1,000 documents retrieved with BM25. 

Next, we describe how we train and evaluate our models on two passage-ranking datasets, MS MARCO and TREC-CAR.

\subsection{Experiments: MS MARCO}

MS MARCO is a passage re-ranking dataset with 8.8M passages obtained from the top-10 results retrieved by the Bing search engine (from 1M queries).
The training set contains approximately 500k pairs of query and relevant documents.
Each query has one relevant passage, on average. The development and test sets contain approximately 6,900 queries each, but relevance labels are made public only for the development set. 

\paragraph{Training}

We fine-tune the model using TPUs\footnote{
\url{https://cloud.google.com/tpu/}
} 
with a batch size of 32 (32 sequences * 512 tokens = 16,384 tokens/batch) for 400k iterations, which takes approximately 70 hours. This corresponds to training on 12.8M (400k * 32) query-passage pairs. We could not see any improvement in the dev set when training for another 10 days, which is equivalent to seeing 50M pairs in total.

We use ADAM~\cite{kingma2014adam} with the initial learning rate set to $3 \times 10^{-6}$, $\beta_1 = 0.9$, $\beta_2 = 0.999$, L2 weight decay of 0.01, learning rate warmup over the first 10,000 steps, and linear decay of the learning rate. We use a dropout probability of $0.1$ on all layers.

\begin{table*}[t]
\begin{center}
\begin{tabular}{l|cc|c}
 & \multicolumn{2}{c|}{MS MARCO} & TREC-CAR\\
 & \multicolumn{2}{c|}{MRR@10} & MAP\\
Method & Dev & Eval & Test\\
\noalign{\vskip 1mm}
\toprule
\noalign{\vskip 1mm}
BM25 (Lucene, no tuning) & 16.7 & 16.5 & 12.3 \\
BM25 (Anserini, tuned) & 18.4 & 18.6 & 15.3 \\
Co-PACRR$^\star$~\cite{macavaney2017contextualized} & - & - & 14.8\\
KNRM~\cite{xiong2017end} & 21.8 & 19.8 & -\\
Conv-KNRM~\cite{dai2018convolutional} & 29.0 & 27.1 & -\\
IRNet$^\dagger$ & 27.8 &  28.1 & -\\
\noalign{\vskip 1mm}
\midrule
\noalign{\vskip 1mm}
BERT Base & 34.7 & - & 31.0\\
BERT Large & 36.5 & 35.8 & 33.5\\
\end{tabular}
\end{center}
\caption{Main Result on the passage re-ranking datasets. $\star$ Best Entry in the TREC-CAR 2017. $\dagger$ Previous SOTA in the MS MARCO leaderboard as of 01/04/2019; unpublished work.}
\label{tab:results_reranking}
\end{table*}

\subsection{Experiments: TREC-CAR}

This is the same dataset described in Section~\ref{sec:qr_experiments}, except that we used different training, validation and test set splits, namely, we use the first four folds for training, the last fold for validation and the automatic annotations benchmark of 2017 as the test set.

\paragraph{Training}

We follow the same procedure described for the MS MARCO dataset to fine-tune our models on TREC-CAR. However, there is an important difference. The official pretrained BERT models\footnote{
\url{https://github.com/google-research/bert}
} 
were pretrained on the full Wikipedia, and therefore they have seen, although in an unsupervised way, Wikipedia documents that are used in the test set of TREC-CAR. Thus, to avoid this leak of test data into training, we pretrained the BERT re-ranker only on the half of Wikipedia used by TREC-CAR's training set.

For the finetuning data, we generate our query-passage pairs by retrieving the top ten passages from the entire TREC-CAR corpus using BM25.\footnote{We use the Anserini toolkit~\cite{Yang:2017:AEU:3077136.3080721,Yang_etal_JDIQ2018} (\url{http://anserini.io/)} to index and retrieve the passages.} This means that we end up with 30M example pairs (3M queries * 10 passages/query) to train our model. We train it for 400k iterations, or 12.8M examples (400k iterations * 32 pairs/batch), which corresponds to only 40\% of the training set. Similarly to MS MARCO experiments, we did not see any gain on the dev set by training the models longer.

\paragraph{Metrics}

\smallskip \noindent To evaluate the effectiveness of the methods on MS MARCO, we use its official metric, mean reciprocal rank of the top-10 documents (MRR@10). For TREC-CAR, we use mean average precision (MAP).

\subsection{Results}

We show the main result in
Table~\ref{tab:results_reranking}. Despite training on a fraction of the data available, the proposed BERT-based models surpass the previous state-of-the-art models by a large margin on both of the tasks.

\begin{figure}
\begin{center}
\centerline{\includegraphics[width=0.7\textwidth]{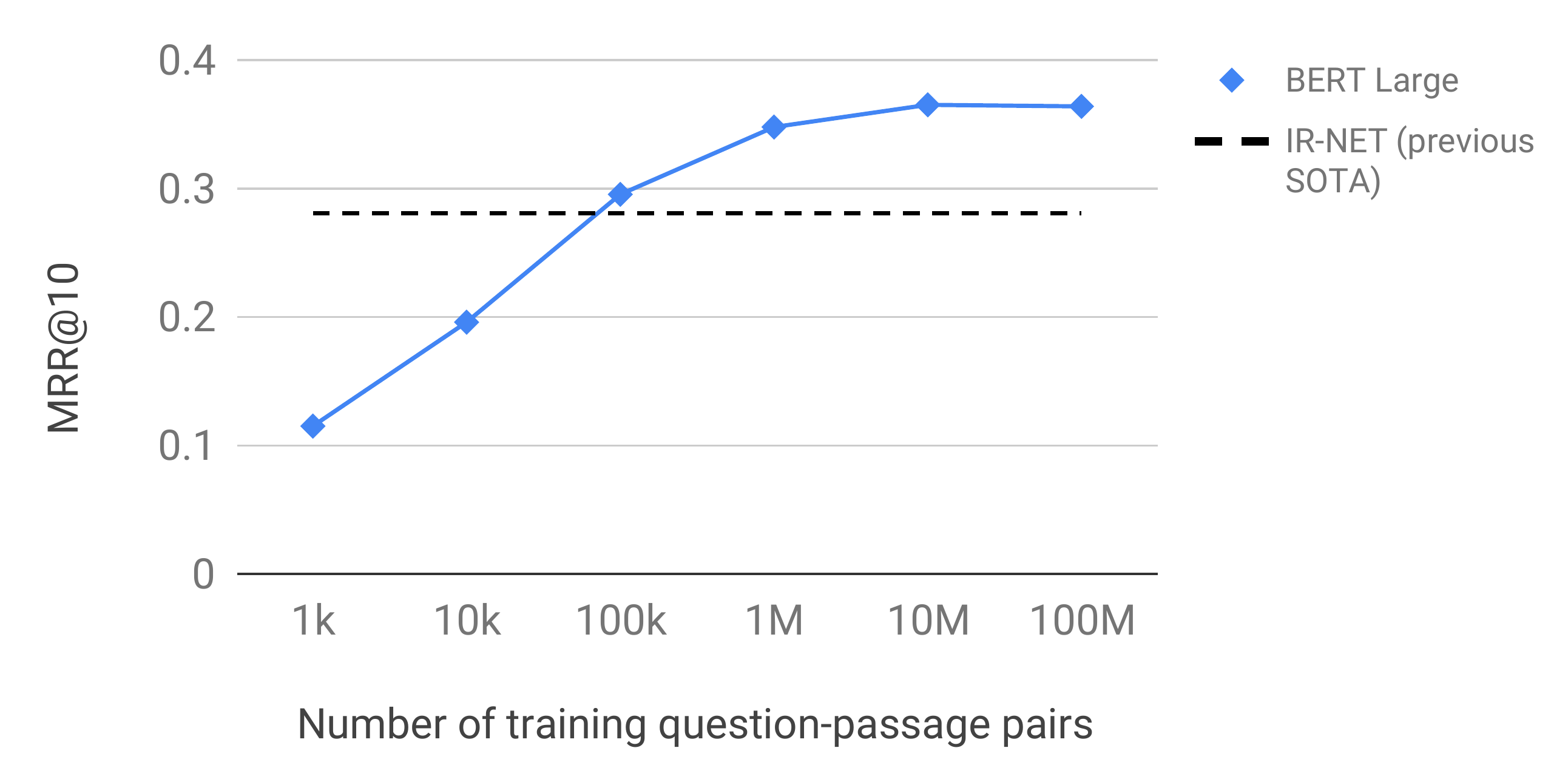}}
\caption{Number of MS MARCO examples seen during training vs. MRR@10.} 
\label{fig:training_size}
\end{center}
\end{figure}

We found that the pretrained models used in this work require a relatively small number of labeled training examples to achieve good effectiveness (Figure~\ref{fig:training_size}). For example, a $\text{BERT}_\text{LARGE}$ trained on 100k question-passage pairs (less than 0.3\% of the MS MARCO training data) is already 1.4 MRR@10 points better than the previous state of the art, IR-NET.

\section{Document Expansion by Predicting Queries}

\begin{figure}
\begin{center}
\centerline{\includegraphics[width=0.75\textwidth]{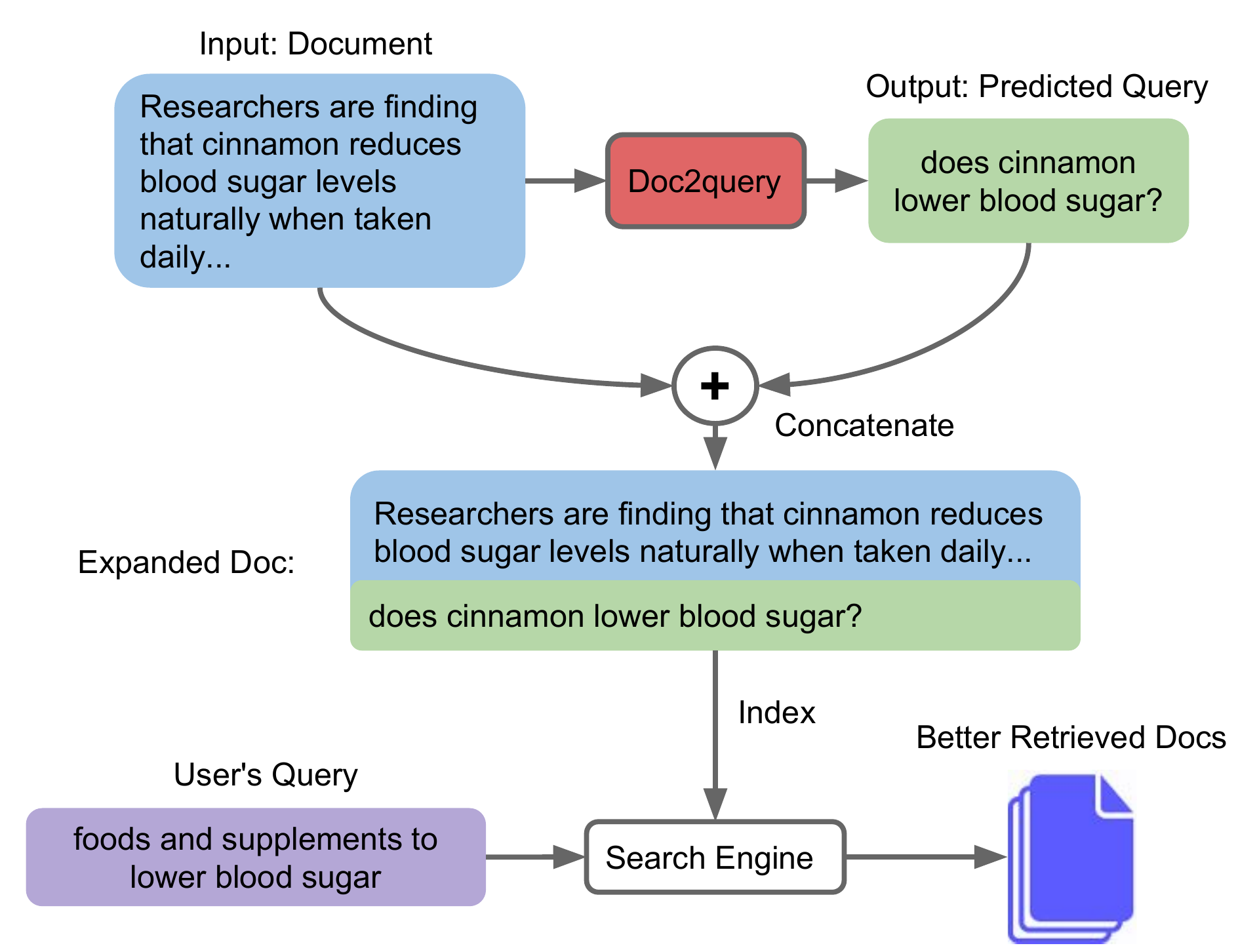}}
\caption{Given a document, our Doc2query model predicts a query, which is appended to the document. Expansion is applied to all documents in the corpus, which are then indexed and searched as before.
}
\label{fig:doc2query_overview}
\end{center}
\end{figure}

Query rewrite is about enriching the query representation while holding the document representation static. Here, we explore an alternative approach based on enriching the document representation (prior to indexing).
Focusing on question answering, we train a sequence-to-sequence model, that given a document, generates possible questions that the document might answer.
An overview of the proposed method is shown in Figure~\ref{fig:doc2query_overview}.

This is the first successful application of document expansion using neural networks.
On the MS MARCO dataset, our approach is competitive to the best results on the official leaderboard (as of 05/31/2019), and we report the best-known results on TREC CAR.
We further show that document expansion is more effective than query expansion on these two datasets.
We accomplish this with relatively simple models using existing open-source toolkits, which allows easy replication of our results.
Document expansion also presents another major advantage, since the enrichment is performed prior to indexing:\
Although retrieved output can be further re-ranked using a neural model to increase effectiveness, the output can also be returned as-is.
These results already yield a noticeable improvement in effectiveness over a ``bag of words'' baseline without the need to apply expensive and slow neural network inference at retrieval time.

\subsection{Related Work}

Prior to the advent of continuous vector space representations and neural ranking models, information retrieval techniques were mostly limited to keyword matching (i.e., ``one-hot'' representations).
Alternatives such as latent semantic indexing~\cite{LSI} and its various successors never really gained significant traction.
Approaches to tackling the vocabulary mismatch problem within these constraints include relevance feedback~\cite{Rocchio_1971}, query expansion~\cite{Voorhees:1994:QEU:188490.188508,Xu00}, and modeling term relationships using statistical translation~\cite{Berger:1999:IRS:312624.312681}.
These techniques share in their focus on enhancing {\it query} representations to better match documents.

In this work, we adopt the alternative approach of enriching {\it document} representations~\cite{tao2006language,pickens2010reverted,efron2012improving}, which works particularly well for speech~\cite{singhal1999document} and multi-lingual retrieval, where terms are noisy.
Document expansion techniques have been less popular with IR researchers because they are less amenable to rapid experimentation.
The corpus needs to be re-indexed every time the expansion technique changes (typically, a costly process); in contrast, manipulations to query representations can happen at retrieval time (and hence are much faster).
The success of document expansion has also been mixed; for example, \citet{billerbeck2005document} explore both query expansion and document expansion in the same framework and conclude that the former is consistently more effective.

A new generation of neural ranking models offer solutions to the vocabulary mismatch problem based on continuous word representations and the ability to learn highly non-linear models of relevance; see recent overviews by~\citet{Onal:2018:NIR:3229901.3229980} and~\citet{MitraBhaskar_Craswell_2019}.
However, due to the size of most corpora and the impracticality of applying inference over every document in response to a query, nearly all implementations today deploy neural networks as re-rankers over initial candidate sets retrieved using standard inverted indexes and a term-based ranking model such as BM25~\cite{robertson1995okapi}.
Our work fits into this broad approach, where we take advantage of neural networks to augment document representations prior to indexing; term-based retrieval then happens exactly as before.
Of course, retrieved results can still be re-ranked by a state-of-the-art neural model, but the output of term-based ranking already appears to be quite good.
In other words, our document expansion approach can leverage neural networks without their high inference-time costs.

\begin{table*}[t]
\centering\centering\resizebox{0.9\textwidth}{!}{
\begin{tabular}{l|ccc|c}
& TREC-CAR & \multicolumn{2}{c|}{MS MARCO} & Retrieval Time\\
& MAP & \multicolumn{2}{c|}{MRR@10} & ms/query\\
& Test & Test & Dev \\
\noalign{\vskip 1mm}
\hline
\noalign{\vskip 1mm}
Single Duet v2~\cite{mitra2019updated} & - & 24.5 & 24.3 &650$^\star$ \\
Co-PACRR$^{\spadesuit}$~\cite{macavaney2017contextualized} & 14.8 & - & - & - \\
\noalign{\vskip 1mm}
\hline
\noalign{\vskip 1mm}
BM25 & 15.3 & 18.6 & 18.4 & 50\\
BM25 + RM3 & 12.7 & - &16.7 & 250 \\
BM25 + Doc2query (Ours) & 18.3 & 21.8 & 21.5 & 90\\
BM25 + Doc2query + RM3 (Ours) &  15.5    & -  & 20.0 & 350\\

\noalign{\vskip 1mm}
\hline
\noalign{\vskip 1mm}
BM25 + BERT & 34.8 & 35.9 & 36.5 & 3400$^{\dagger}$\\
BM25 + Doc2query + BERT (Ours) & \textbf{36.5} & \textbf{36.8} & \textbf{37.5} & 3500$^{\dagger}$\\
\end{tabular}}
\caption{Main results on TREC-CAR and MS MARCO datasets. $^\star$ Our measurement, in which Duet v2 takes 600ms per query, and BM25 retrieval takes 50ms. $^{\spadesuit}$ Best submission of TREC-CAR 2017. $^{\dagger}$ We use Google's TPUs to re-rank with BERT. In bold are statistically significant results ($p < 0.05$) according to Student's paired t-test with a Bonferroni correction (code modified from \url{https://github.com/castorini/Anserini/blob/master/src/main/python/compare_runs.py}) }
\label{tab:main_results}
\end{table*}

\subsection{Method}

Our proposed method, which we call ``Doc2query'', proceeds as follows:\
For each document, the task is to predict a set of queries for which that document will be relevant.
Given a dataset of query-relevant document pairs, we use a sequence-to-sequence Transformer model~\cite{vaswani2017attention} that takes as an input the document terms and produces a query.
The document and target query are segmented using BPE~\cite{sennrich2015neural} after being tokenized with the Moses tokenizer.\footnote{
\url{http://www.statmt.org/moses/}
} 
To avoid excessive memory usage, we truncate each document to 400 tokens and queries to 100 tokens.

The architecture of our transformer model is identical to the \textit{base} model described in \citet{vaswani2017attention}, which has 6 layers for both encoder and decoder, 512 hidden units in each layer, 8 attention heads and 2048 hidden units in the feed-forward layers.
We train with a batch size of 4096 tokens for a maximum of 30 epochs. 
We use Adam with a learning rate of $10^{-3}$, $\beta_1$ = 0.9, $\beta_2$ = 0.998, L2 weight decay of 0.01, learning rate warmup over the first 8,000 steps, and linear decay of the learning rate. We use a dropout probability of 0.1 in all layers. 
Our implementation uses the OpenNMT framework~\cite{opennmt}; training takes place on four V100 GPUs.
To avoid overfitting, we monitor the BLEU scores of the training and development sets and stop training when their difference is larger than four points. 

Once the model is trained, we predict 10 queries using top-$k$ random sampling~\cite{fan2018hierarchical} and append them to each document in the corpus. 
We do not put any special markup to distinguish the original document text from the predicted queries. 
The expanded documents are indexed, and we retrieve a ranked list of documents for each query using BM25. 
We optionally re-rank these retrieved documents using BERT as described in Section~\ref{sec:reranker}.

\subsection{Baselines}

\smallskip \noindent We evaluate the following methods:

\smallskip \noindent {\bf BM25}:\ We use the Anserini open-source IR toolkit
to index the original (non-expanded) documents and BM25 to rank the passages. During evaluation, we use the top-1000 re-ranked passages.

\smallskip \noindent {\bf BM25 + Doc2query}:\ We first expand the documents using the proposed Doc2query method. We then index and rank the expanded documents exactly as in the BM25 method above.

\smallskip \noindent {\bf RM3}: To compare document expansion with query expansion, we applied the RM3 query expansion technique~\cite{Abdul-Jaleel04}.
We apply query expansion to both unexpanded documents (BM25 + RM3) as well as the expanded documents (BM25 + Doc2query + RM3).

\smallskip \noindent {\bf BM25 + BERT}:\ We index, and retrieve documents as in BM25 and further re-rank the documents with the BERT Large described in Section~\ref{sec:reranker}.

\smallskip \noindent {\bf BM25 + Doc2query + BERT}:\ We expand, index, and retrieve documents as in BM25 + Doc2query and further re-rank the documents with BERT.

\begin{table*}
\begin{center}
\begin{small}
\begin{tabular}{ll}
\noalign{\vskip 1mm}
\hline
\noalign{\vskip 1mm}
Input Document: & July is the hottest month in Washington DC with an average\\
& temperature of 27C (80F) and the coldest is January at 4C (38F) \\
& with the most daily sunshine hours at 9 in July. The wettest \\
& month is May with an average of 100mm of rain.\\
Predicted Query: & weather in washington dc\\
Target query: & what is the temperature in washington\\
\noalign{\vskip 1mm}
\hline
\noalign{\vskip 1mm}
Input Document: & The Delaware River flows through Philadelphia into the Delaware\\
& Bay. It flows through and aqueduct in the Roundout Reservoir\\
& and then flows through Philadelphia and New Jersey before\\
& emptying into the Delaware Bay.\\
Predicted Query: & what river flows through delaware\\
Target Query: & where does the delaware river start and end\\
\noalign{\vskip 1mm}
\hline
\noalign{\vskip 1mm}
Input Document: & sex chromosome - (genetics) a chromosome that determines the\\
& sex of an individual; mammals normally have two sex\\
& chromosomes chromosome - a threadlike strand of DNA in the\\
& cell nucleus that carries the genes in a linear order; humans have\\
& 22 chromosome pairs plus two sex chromosomes.\\
Predicted Query: & what is the relationship between genes and chromosomes\\
Target Query: & which chromosome controls sex characteristics\\
\noalign{\vskip 1mm}
\hline
\noalign{\vskip 1mm}
\end{tabular}
\end{small}
\end{center}
\vspace{-4mm}
\caption{Examples of query predictions on MS MARCO compared to real user queries.
}
\label{tab:examples}
\end{table*}

\subsection{Results}

Results on both datasets are shown in Table~\ref{tab:main_results}.
BM25 is the baseline.
Document expansion with our method (BM25 + Doc2query) improves retrieval effectiveness by $\sim$15\% for both datasets.
When we combine document expansion with a state-of-the-art re-ranker (BM25 + Doc2query + BERT), we achieve the best-known results to date on TREC CAR; for MS MARCO, we are near the state of the art.
Our full re-ranking condition (BM25 + Doc2query + BERT) beats BM25 + BERT alone, which verifies that the contribution of Doc2query is indeed orthogonal to that from post-indexing re-ranking.

\subsection{Evaluating Various Decoding Schemes}

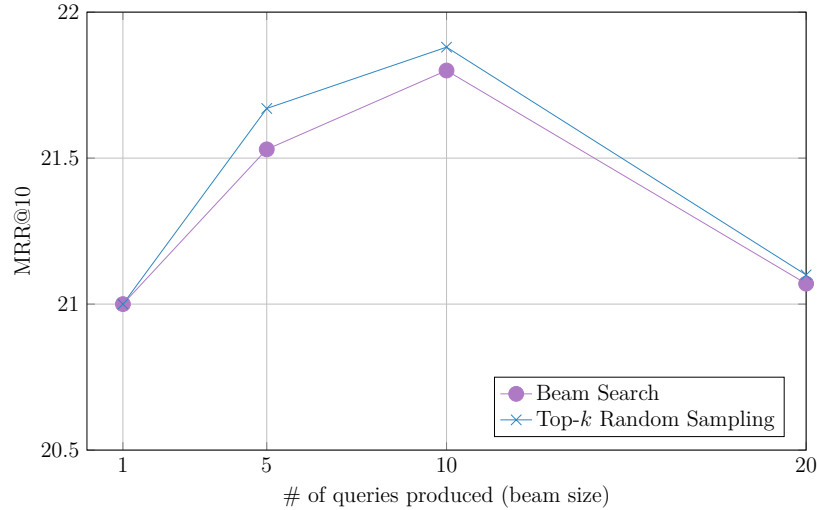
\begin{figure}[t]
\centering
\begin{tikzpicture}[scale = 0.7]
\begin{axis}[
width=1.0\columnwidth,
height=0.65\columnwidth,
legend cell align=left,
legend style={at={(1, -0.1)},anchor=south east,font=\normalsize},
mark options={mark size=3},
font=\normalsize,
xmin=0, xmax=20,
ymin=20.5, ymax=22,
xtick={1, 5, 10, 20},
ytick={20.5, 21.0, 21.5, 22},
legend pos=south east,
xmajorgrids=true,
ymajorgrids=true,
xlabel style={font = \normalsize, yshift=0ex},
xlabel=\# of queries produced (beam size),
ylabel=MRR@10,
ylabel style={font = \normalsize, yshift=0ex}
]    legend entries={Recall,
        Top 1 EM},
    ]
    \addplot[mark=*,g-purple, mark options={scale=2}] plot coordinates {
   (1, 21)(5, 21.53)(10, 21.8)(20, 21.07)
    };
    \addlegendentry{Beam Search}
	\addplot[mark=x,g-blue, mark options={scale=2}] plot coordinates {
	   (1, 21)(5, 21.67)(10, 21.88)(20, 21.1)
	    };
    \addlegendentry{Top-$k$ Random Sampling}
    \end{axis}
    
    \end{tikzpicture}
\caption{Retrieval effectiveness on the development set of MS MARCO when using different decoding methods to produce queries. On the {\it x}-axis, we vary the number of predicted queries that are appended to the original documents.} 
\label{fig:decoding_comparison}
\vspace{0mm}
\end{figure}

Here we investigate how different decoding schemes used to produce queries affect the retrieval effectiveness.
We experiment with two decoding methods:\ beam search and top-$k$ random sampling with different beam sizes (number of generated hypotheses). Results are shown in Figure~\ref{fig:decoding_comparison}.
Top-$k$ random sampling is slightly better than beam search across all beam sizes, and we observed a peak in the retrieval effectiveness when 10 queries are appended to the document.
We conjecture that this peak occurs because too few queries yield insufficient diversity (fewer semantic matches) while too many queries introduce noise and reduce the contributions of the original text to the document representation.

\subsection{Qualitative Analysis}

Where exactly are these better scores coming from?
We show in Table~\ref{tab:examples} examples of queries produced by our Doc2query model trained on MS MARCO.
We notice that the model tends to copy some words from the input document (e.g., Washington DC, River, chromosome), meaning that it can effectively perform term re-weighting (i.e., increasing the importance of key terms).
Nevertheless, the model also produces words not present in the input document (e.g., weather, relationship), which can be characterized as expansion by synonyms and other related terms.

To quantify this analysis, we measured the proportion of predicted words that exist (copied) vs. not-exist (new) in the original document.
Excluding stop words, which corresponds to 51\% of the predicted query words, we found that 31\% are new while the rest (69\%) are copied.
If we expand MS MARCO documents using only new words and retrieve the dev set queries with BM25, we obtain an MRR@10 of 18.8 (as opposed to 18.4 when indexing with original documents).
Expanding with copied words gives an MRR@10 of 19.7.
We achieve a higher MRR@10 of 21.5 when documents are expanded with both types of words, showing that they are complementary.

Further analyses show that one source of improvement comes from having more relevant documents for the re-ranker to consider.
We find that the Recall@1000 of the MS MARCO dev set increased from 85.3 (BM25) to 89.3 (BM25 + Doc2query).
Results show that BERT is indeed able to identify these correct answers from the improved candidate pool and bring them to the top of the ranked list, thus improving the overall MRR.

As a contrastive condition, we find that query expansion with RM3 hurts in both datasets, whether applied to the unexpanded corpus (BM25 + RM3) or the expanded version (BM25 + Doc2query + RM3).
This is somewhat a surprising result because query expansion usually improves effectiveness in document retrieval, but this can likely be explained by the fact that both MS MARCO and CAR are precision oriented.
This result shows that document expansion can be more effective than query expansion, most likely because there are more signals to exploit as documents are much longer.

Finally, for production retrieval systems, latency is often an important factor.
Our method without a re-ranker (BM25 + Doc2query) adds a small latency increase over baseline BM25 (50 ms vs.\ 90 ms) but is approximately seven times faster than a neural re-ranker that has a three points higher MRR@10 (Single Duet v2, which is presented as a baseline in MS MARCO by the organizers).
For certain operating scenarios, this tradeoff in quality for speed might be worthwhile.

\section{Summary}

In this chapter, we introduced two novel components of a search engine.
One is a simple adaptation of BERT as a passage re-ranker, which showed to be very effective to increase precision. 
The other component aims to improve recall by expanding documents prior to indexing.
It is the first successful use of document expansion based on neural networks.
Furthermore, this approach allows developers to shift the computational costs of neural network inference from retrieval to indexing.

A commonly overlooked issue when independently tuning multiple components of a pipeline is that their individual improvements might not be additive when they run together~\cite{armstrong2009improvements,sculley2018winner,lin2019neural}. This was not the case here: we showed that the gains from our two novel components are orthogonal, and we were able to build a state-of-the-art retrieval system when we combined them.

Our implementation is based on integrating three open-source toolkits:\ OpenNMT, Anserini, and TensorFlow BERT.
The relative simplicity of our approach aids in the reproducibility of our results and paves the way for further improvements in re-ranking and document expansion.

\chapter{Conclusion}
\label{chap:conclusion}

Our main motivation for this thesis was to build machines that can produce answers based on pieces of information found in a large corpus. While investigating possible solutions, we realized that it is very challenging to select relevant data to support the answer. This difficulty is, in part, due to the vast amount of unstructured information available, the majority of it is either irrelevant or provide incorrect facts. As evidence that the problem is indeed hard, some claim that there has been little progress over the past two decades in creating better retrieval systems~\cite{armstrong2009improvements,lin2019neural}. These observations led us to work on mechanisms that can more effectively retrieve relevant information to a given question, i.e., search engines.

We first showed that it is possible to build a search engine that operates in a very different way from traditional ones by training an agent that finds documents by navigating the web via hyperlinks. We then showed that it is possible to train agents that learn how to retrieve better documents from the search engine, treated as a black box, by rewriting queries. Lastly, we focused on improving the internal components of the search engine, namely, the re-ranker and the inverted index, which resulted in a retrieval system with twice the performance of an off-the-shelf search engine~\cite{Yang:2017:AEU:3077136.3080721,Yang_etal_JDIQ2018}.

\paragraph{Future Work:} With better retrieval methods, we hope that future machines will evolve to a kind of oracle, that is, a machine that can answer our questions, no matter how difficult they are, and help us expand our understanding of the world.

To create such a system, we argue that we need harder question-answering datasets that are motivated by real use cases. One instance of such dataset could be: given a real user question whose answer is unlikely to be found in any single document of a large corpus, the task would be to generate an answer in a natural language whose supporting evidence comes from multiple documents. This answer is, in essence, a summary of multiple documents based on a question. The answer could be a single sentence, a table, or even an entire article. Hence, if existing search engines find a few bits of information in a large collection of documents, the next generation of these systems will connect the bits and surface patterns in a form that is easy to ingest by a human reader. 

A system that performs well in this task is a good candidate to help, for example, biomedical scientists in finding which genes are related to which diseases and, in the process, it might also suggest gene-disease relations there were not explored in the literature. Similarly, it can assist doctors in recommending drugs to a patient, given her previous and current conditions. For that, the system would have to consider the effects of each component of the drug in the patient, which can be facilitated if the system has access to a history of treatments of patients that share similar conditions with the current patient.



\begin{singlespace}
\bibliographystyle{plainnat}
\bibliography{main}
\end{singlespace}

\begin{appendices}
\chapter{Reproducibility}

The code and datasets to reproduce our results are available at:

\begin{itemize}
\item Chapter~\ref{chap:webnav}: \url{https://github.com/nyu-dl/dl4ir-webnav}

\item Chapter~\ref{chap:query_reformulation}: \url{https://github.com/nyu-dl/dl4ir-query-reformulator}

\item Chapter~\ref{chap:blackbox}, Re-ranker: \url{https://github.com/nyu-dl/dl4marco-bert}

\item Chapter~\ref{chap:blackbox}, Doc2query: \url{https://github.com/nyu-dl/dl4ir-doc2query}

\end{itemize}

\end{appendices}
\end{document}